\documentclass[11pt,a4paper]{article}

\usepackage{amssymb,amsmath,amsfonts}
\usepackage{graphicx}
\usepackage{color}
\usepackage{ulem}             
\usepackage{multicol}
\usepackage{caption}
\usepackage{pgf,tikz}
\usetikzlibrary{arrows}
\usepackage{lipsum}
\newcommand\blfootnote[1]{%
  \begingroup
  \renewcommand\thefootnote{}\footnote{#1}%
  \addtocounter{footnote}{-1}%
  \endgroup
}
	\usetikzlibrary[patterns]

\usepackage{natbib}
\newcommand{\figpath}{./}

\newcommand\lam{{\lambda} }

\newcommand{\RR}{{\mathbb R}}

\newcommand\ab{{\overline a} }

\newcommand\gO{{\cal O} }

\newcommand\gH{{\cal H} }

\newcommand{\be}{\begin{equation}}
\newcommand{\ee}{\end{equation}}

\newcommand{\tpr}[1]{{#1}}

\newenvironment{disarray}%
 {\everymath{\displaystyle\everymath{}}\array}%
 {\endarray}

\begin{document}

\title{On the rotation of co-orbital bodies in eccentric orbits}

\author{A. Leleu\footnote{IMCCE, Observatoire de Paris - PSL Research University, UPMC Univ. Paris 06, Univ. Lille 1, CNRS, 77 Avenue Denfert-Rochereau, 75014 Paris, France} \and  P. Robutel$^*$ \and  A.C.M. Correia$^*$\footnote{Departemento de F\`isica, I3N, Universidade de Aviero, Campus de Santiago, 2810-193 Aveiro - Portugal}
                 }     

\maketitle

\begin{abstract}

We\blfootnote{E-mail: adrien.leleu@obspm.fr; philippe.robutel@obspm.fr  and correia@ua.pt} investigate the resonant rotation of co-orbital bodies in eccentric and planar orbits. We develop a simple analytical model to study the impact of the eccentricity and orbital perturbations on the spin dynamics. This model is relevant in the entire domain of horseshoe and tadpole orbit, for moderate eccentricities. We show that there are three different families of spin-orbit resonances, one depending on the eccentricity, one depending on the orbital libration frequency, and another depending on the pericenter's dynamics. We can estimate the width and the location of the different resonant islands in the phase space, predicting which are the more likely to capture the spin of the rotating body. In some regions of the phase space the resonant islands may overlap, giving rise to chaotic rotation.


\end{abstract}

\section{Introduction}
\label{sec:intro}

In 1772, Lagrange has found an equilibrium configuration where three bodies are located at the vertices of an equilateral triangle where they all move with the same orbital period. 
\citet{1843GG} has proved the stability of this configuration in the case of circular motion, providing a stability criteria for the masses of the bodies. 
There are two stable configurations for a quasi-circular co-orbital system: the tadpole orbits, where the two bodies librate around the Lagrangian equilibrium; and the horseshoe orbits, named after the shape the trajectories of the bodies describe in the rotating frame. These configurations are stable if the ratio between the sum of the masses of the co-orbitals and the total mass of the system is below $1/27$ for tadpole orbit \citep{1843GG} and $ \approx 2 \times 10^{-4}$ for horseshoe orbits \citep[see for example][]{LeRoCo2015}. For eccentric co-orbitals, the previous configurations continue to exist, but more configurations are possible: quasi-satellite orbits, in which the two bodies appear to have a retrograde orbit around each other \citep{Namouni1999,Mik2006}; retrograde co-orbitals, where the co-orbitals orbit in opposite direction \citep{MoNa2013}; and anti-Lagrange orbits, which, with the eccentric Lagrangian equilibrium, correspond to the two planar Lyapunov families of orbit emanating from the circular Lagrangian equilibrium \citep{GiuBeMiFe2010,RoPo2013}. 
The first body on a tadpole orbit was found at the $L_4$ point of Jupiter \citep{Wolf1906} and we presently know more than 6000 bodies of this kind in the solar system, in the frame of the restricted three body problem\footnote{http://www.minorplanetcenter.org/}.
Bodies in the horseshoe configuration have been discovered around Saturn, where the co-orbitals have commensurable masses \citep{DeMu1981a}. 
Until now, no long-term stable example of the three other co-orbitals configuration have been found.

For close-in bodies, tidal interactions slowly modify the rotation and the orbits \citep[e.g.][]{1964MG,2009C}. 
When the rotation rate and the mean motion have the same magnitude, the dissipative tidal torque may be counterbalanced by the conservative torque due to the axial asymmetry of the inertia ellipsoid. In the two-body problem, for circular orbits, the only possibility for the spin is to end at the synchronous resonance \citep{1966GP,2009CL}. 
However, for eccentric orbits, the rotation rate can be locked in a half-integer commensurability with the mean motion, usually called spin-orbit resonance \citep{1965C,1966GP,2009CL} or have a chaotic rotation, as it is the case for Hyperion \citep{WiPeMi1984}. In the co-orbital circular case, the presence of a co-orbital companion can also give rise to non-synchronous spin-orbit resonances due to the orbital libration around the Lagrangian point (see \cite{2013CR}). In this paper we develop an analytic model for the planar rotation which can take into account both the effect of the eccentricity of the rotating body and the perturbation by a co-orbital. By setting the mass of the co-orbital companion to zero, we retrieve the results of \cite{1965C} for a Keplerian eccentric orbit. By setting the eccentricity to zero we retrieve the results of \cite{2013CR} for circular co-orbitals. We then study the rotational dynamics for eccentric co-orbitals. The case of fixed value of the eccentricities and the longitudes of the perihelion is studied section \ref{sec:CoC}, while the effect of the variation of these parameters is described in section \ref{sec:evar}.

\section{The co-orbital dynamics}
\label{sec:orb}
\subsection{Equation of motion of the co-orbitals}

\tpr{We denote $m_0$ the mass of the central body, and $m_1$ and $m_2$ the masses of the co-orbital bodies.} At order one in the eccentricity $e$, The variable $\zeta = \lambda_1 - \lambda_2$ satisfies the second order differential equation: 

\begin{equation}
\ddot\zeta=-3\mu n^2\left[1-(2-2\cos\zeta)^{-3/2}\right]\sin\zeta\ ,
\label{eq:sol_orb_zeta} 
\end{equation}%

with 
\begin{equation}
\mu=\frac{m_1+m_2}{m_0+m_1+m_2} \ .
\end{equation} 
\begin{figure}[h!]
\begin{center}
\includegraphics[width=.7\linewidth]{\figpath/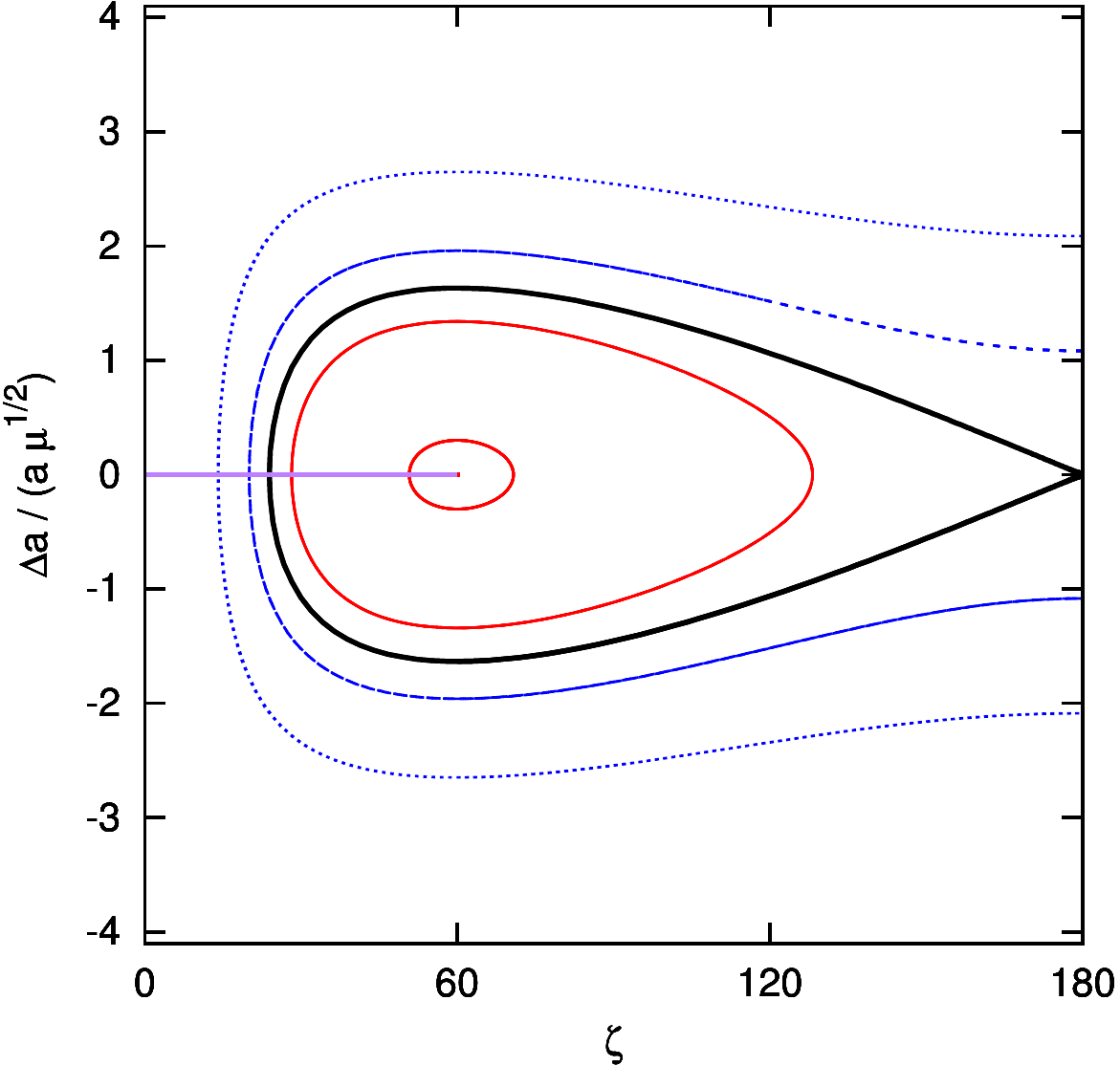}
\caption{\label{fig:orbit} Phase portrait  of  equation (\ref{eq:sol_orb_zeta}). The vertical axis is equivalent to the variable $u$ defined by \cite{RoPo2013}, at the order $1$ in the masses. The separatrix (black curve) splits the phase space in two different domains: inside the separatrix the region associated with the tadpole orbits (in red) and the horseshoe domain (in blue) outside. The phase portrait is symmetric with respect to $\zeta=180^\circ$.
The horizontal purple segment indicates the range of variation of $\zeta_0$. See the text for more details.}
\end{center}
\end{figure}
This differential equation is one of the most common representations of the coorbital motion \citep[see][and references therein]{Morais1999,RoNi2015}. It describes the relative motion of the two bodies and it is valid as long as the coorbital bodies are not too close to the collision ($\zeta=0$).  
Since equation (\ref{eq:sol_orb_zeta}) is invariant under the symmetry $\zeta\longmapsto 2\pi-\zeta$, the study of its phase portrait can be reduced to the domain $(\zeta,\dot\zeta) \in [0, \pi]\times \RR$  (see Fig. \ref{fig:orbit}). The equilibrium point located at $(\zeta,\dot\zeta) = (\pi/3,0)$ corresponds  to one of the two Lagrangian equilateral configurations\footnote{the coordinates of the other point are $(5\pi/3,0)$. The permutation of the index $1$ and $2$ of the planets allows to exchange the two equilateral configurations, which are linearly stable for small enough  planetary masses \citep[namely, if $\frac{m_0 m_1+m_1 m_2+m_0 m_2} {(m_0+m_1+m_2)^2} < \frac{1}{27} \approx 0.037$, see][]{1843GG}.}. In the vicinity of this equilibrium, at order one in $(\zeta-\pi/3)$, the frequency of the motion is close to \citep{Charlier1906}:
\be
\nu= n \sqrt{\frac{27}{4} \mu}.
\label{eq:nu}
\ee
The other equilibrium, whose coordinates are $(\pi,0)$, corresponds to the unstable Eulerian collinear configuration of the type $L_3$. The separatrices (in black in Figure \ref{fig:orbit}) emanating from this last unstable point divide the phase space in three different regions: two corresponding to the tadpole trajectories surrounding one of the two Lagrange's equilibria (in red), and another corresponding to the horseshoe orbits, which encompasses the three above-mentioned fixed points (in blue).
As shown in Fig. \ref{fig:orbit}, any trajectory given by equation (\ref{eq:sol_orb_zeta}) can be entirely determined by the initial conditions $(t_0,\zeta_0)$ such that $\zeta(t_0)=\zeta_0$ and $\dot\zeta(t_0) =0$, where $\zeta_0$ is the minimum value of $\zeta$ along the trajectory, and $t_0$ the first positive instant for which the value $\zeta_0$ is reached. 

%
%
The possible values of $\zeta_0$, represented by the purple horizontal line on Figure \ref{fig:orbit}, are included in the interval $(0^\circ,60^\circ]$. $\zeta_0 = 60^\circ$ corresponds to the equilateral configuration where $m_1$ is the leading body and $m_2$ the trailing one. 
The tadpole orbits correspond to $\zeta_0 \in (\zeta_s, 60^\circ]$, where $\zeta_s \approx 23.9^\circ$  corresponds to the separatrix, while $\zeta_0$ ranges from $\zeta_s$ to $0$ for horseshoe orbits.

The expressions of $\lam_j$ and $a_j$, the semi-major axis of the body $j$, are given by \citep{RoNi2015}:
\be
\lam_j = \frac{z_0}{2} + nt + (-1)^{j+1} \frac{m_k}{m_1+m_2}\zeta +\gO(\mu, e\sqrt{\mu}, e^2), \quad k\neq j
\label{eq:lam_sol}
\ee
 and 
\be
a_j = \ab\left(1  + (-1)^j\frac23 \frac{m_k}{m_1+m_2}\frac{\dot\zeta}{n}\right) +\gO(\mu, e\mu, e^2), \quad k\neq j,
\label{eq:a_sol}
\ee
 where $z_0$ is a constant that is determined by the value of $\lambda_j$ at $t=0$. Using the usual expansions in eccentricity power of the distance $r_j$ between the star and the planet $j$ and of its true longitude $f_j$ given by:
\be
r_j = a_j\left(1 - e_j\cos(\lam_j -\varpi_j) \right) +\gO(e^2), \quad
f_j = \lam_j + 2e_j\sin(\lam_j -\varpi_j)  +\gO(e^2),
\ee
and injecting (\ref{eq:lam_sol}) and (\ref{eq:a_sol}) in these formulas, we get the expressions:
\be
\begin{split}
&r_j = \ab\left(1 - e_j\cos(\lam_j -\varpi_j) + (-1)^j\frac23 \frac{m_k}{m_1+m_2}\frac{\dot\zeta}{n} \right) +\gO(\mu,e\mu,e^2), \\
&f_j = \lam_j + 2e_j\sin(\lam_j -\varpi_j)  +\gO(\mu,e\sqrt{\mu},e^2),
\end{split}
\label{eq:polar}
\ee
where $\lam_j$ is given by (\ref{eq:lam_sol}) and $\zeta$ satisfies the differential equation (\ref{eq:sol_orb_zeta}). 
In expressions (\ref{eq:polar}), the eccentricities and the longitudes of the periastron are not necessarily constant, but their variations occur on a much longer time-scale than the one associated to the variations of $\zeta$. 

According to \cite{RoPo2013}, in the vicinity of $L_4$ or $L_5$, the temporal variations of $X_j = e_j\exp(i\varpi_j)$ can be approximated by the expressions 
\be
X_1(t) =  \operatorname{e}^{i\zeta_L}\left(
\sqrt{\frac{m_2}{m_1}} z_1 \operatorname{e}^{ig_1 t}  + z_2 \operatorname{e}^{ig_2 t} 
\right), \quad
X_2(t) =  
- \sqrt{\frac{m_1}{m_2}} z_1 \operatorname{e}^{ig_1 t}  + z_2 \operatorname{e}^{ig_2 t} .
\label{eq:secular_sol}
\ee
where $\zeta_L = \pi/3$ ou $-\pi/3$ depending on the selected Lagrange configuration. The complex numbers $z_1$ and $z_2$ are two constants that can be determined by the initial values of $X_1$ and $X_2$, and $g_1$ and $g_2$ are the two eigenfrequencies of the differential system associated to the $X_i$. If we introduce the quantities $\rho_1, \rho_2$ and $\varphi$ such that $z_1 = \rho_1$ and $z_2 = \rho_2 \operatorname{e}^{i\varphi}$, the elliptic Lagrange family arises for $\rho_1=0$ and $\rho_2>0$. The two eccentricities are equal and the apsidal lines are fixed with $\varpi_1 -\varpi_2 = \zeta_L$. 

At the Lagrangian points one of the eigenfrequencies vanish, since these equilibria points are degenerated. More precisely, we have:
\be 
  g_1 =  \frac{27}{8}\frac{m_1+m_2}{m_0} n, \quad g_2 = 0.
\ee
 It is worth mentioning  that the degeneracy is removed as soon as one moves away from the equilibrium point. Indeed, in the neighbourhood of the Lagrangian equilibrium, it can be shown that \citep{RoPo2013}:
 \be
 g_2 = \gO\left( \frac{m_1+m_2}{m_0} {\max}(\zeta - \zeta_L)^2 \right).
 \ee
Consequently, we are left with 4 different time-scales: $2\pi/n = \gO(1)$ associated to the orbital motions, $2\pi/\nu = \gO(1/\sqrt{\mu})$ which governs the variations of $\zeta$ and of the semi-major axis $a_j$, $2\pi/g_1 = \gO(1/\mu)$ corresponding to the secular variations of the eccentricities and $\varpi_j$, and $2\pi/g_2$ which is associated to the precession of the perihelion and which is the slowest one \citep[we always have $ g_2 \ll g_1$,][]{RoPo2013}. In the vicinity of this configuration, for $0<\rho_1 \ll \rho_2$, the temporal variations of $e_j$ and $\varpi_j$ can be approximated by:
\be
\begin{split}
e_j &= \rho_2\left( 
  1+ (-1)^{j+1}\sqrt{\frac{m_k}{m_j}}  \frac{\rho_1}{\rho_2} \cos(g t - \varphi) + \gO_2\left(\frac{\rho_1}{\rho_2}\right)
  \right) ,  \\
\varpi_j &\approx \delta_{j,1}\zeta_L +\varphi  + g_2t-  (-1)^{j+1}\sqrt{\frac{m_k}{m_j}}\frac{\rho_1}{\rho_2}\sin(g t - \varphi)
+ \gO_2\left(\frac{\rho_1}{\rho_2}\right),
\end{split}
\label{eq:ew}
\ee  
where $g = g_1-g_2 \approx g_1$, and $\delta_{j,1} = 1$ if  $j=1$ and $0$  otherwise. The two eccentricities are almost equal undergoing small periodic variations at the frequency $g$ in opposition of phase.  The two apsidal lines precess at a very low common frequency (and are fixed if $\zeta = \zeta_L$), and librate at the frequency $g$ with an amplitude of $(\sqrt{m_1/m_2} + \sqrt{m_2/m_1})\rho_1\rho_2$.

\section{Spin Dynamics}

\label{sec:sor}
\begin{figure}[h!]
\begin{center}
\includegraphics[width=.6\linewidth]{\figpath/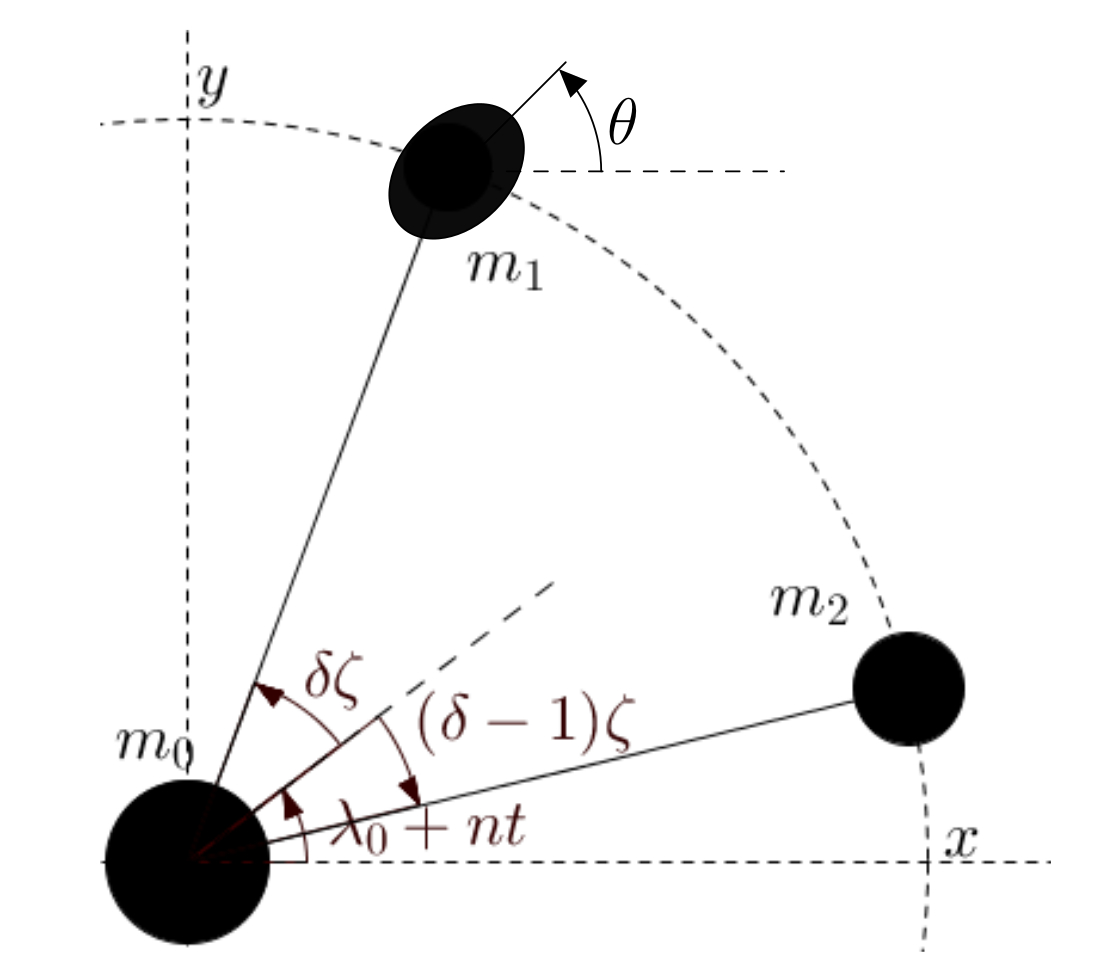}
\caption{\label{fig:angles} Reference angles represented for the circular coorbital system with respect to an inertial frame $(x,y)$. $m_0$ is the mass of the central star, $m_1$ and $m_2$ the mass of the coorbitals. $r_i$ is the distance of the coorbital $i$ to the central star, and $\lambda_i$ its true longitude. $n$ is the mean motion of the planets' barycentre and $\zeta=\lambda_1-\lambda_2$. Following equation (\ref{eq:lam_sol}) one can write $\lambda_i$ as a function of $\lambda_0$, $\zeta$ and the mass ratio $\delta = \frac{m_2}{m_1+m_2}$. $\theta$ is the rotation angle of $m_1$, defined with respect to an inertial frame, that will be used in the following sections.}
\end{center}
\end{figure}

We focus our study in the rotation of the body $1$, since the co-orbital problem is symmetric.
Thus, hereafter we drop the indices $1$ ($r_1$ becomes $r$ etc.). 
We denote $A < B < C$ the moments of inertia of the rotating body. 
For simplicity, we assume that the body is rotating around its main inertia axis which is held perpendicular to the orbital plane. 
Therefore, the rotation of the body can be described by the rotation angle $\theta$ (see Fig.~\ref{fig:angles}). 
The equation of motion for $\theta$ is given by the canonical equations associated with the Hamiltonian $\gH$ given by \citep[e.g.][]{dan1964}
\begin{equation}
\label{eq:thepp} 
\gH=T +\frac{I^2}{2} - \frac{\sigma^2}{4} \left( \frac{\bar{a}}{r}(t) \right)^3 \cos\ 2(\theta-f(t)),
\end{equation}
where $f$ is the true longitude and
\begin{equation}
\label{eq:sig1}
\sigma= n \sqrt{3 \frac{B-A}{C}} 
\end{equation}
is a measure of the axial asymmetry of the rotating body. This parameter is considered as constant in this study (rigid body). $(B-A)/C$ for bodies of the solar system can be found in Appendix \ref{sec:bmasc}.

For circular unperturbed Keplerian orbits, $r=\bar{a}$ and $f = nt$, so the Hamiltonian (\ref{eq:thepp}) is equivalent to the equation of a simple pendulum and it is consequently integrable. The single resonant island is centred on $\dot{\theta}=n$ and the maximum width of this island (in the direction of $\dot{\theta}$) is $2\sigma$. This resonance is called the synchronous spin-orbit resonance.

In the co-orbital eccentric case, the equations of motion are not integrable. 
However, an eccentric orbit perturbed by a co-orbital companion remains quasi-periodic. As a consequence, the elliptic elements of the body can be expanded in a Fourier series whose frequencies are the fundamental frequencies of the orbit ($n$, $\nu$, $g$ and $g_2$), see section \ref{sec:orb}. Following the D'Alembert rules, the expansion of the time-dependent quantity $ \left( \frac{a}{r} \right)^3 \operatorname{e}^{i 2 f}$ that appears in equation (\ref{eq:thepp}) can be written as:
\be 
  \left( \frac{\bar{a}}{r} \right)^3 \operatorname{e}^{i 2 f} = \sum_{j \geq 0 } \ \hat{\rho}^2_{\eta_j} \ \operatorname{e}^{i( 2 \langle \eta_j, \varsigma \rangle t+ \phi_j )}.
\label{eq:asreqp}
\ee
where $\varsigma=(n-g_2,\nu,g) \in  \mathbb{R}^3_+$, $\eta_j=(p,q,s)$ with $2\eta_j \in \mathbb{Z}^3$, $ \langle \eta_j, \varsigma \rangle = p(n-g_2)+ q \nu+ s g$, and $\hat{\rho}_{\eta_j} \in \mathbb{R}_+$ with:  
\be
\hat{\rho}^2_{\eta_j} \operatorname{e}^{i \phi_j} = \frac{1}{\pi} \int_0^{\pi} \left( \frac{\bar{a}}{r} \right)^3 \operatorname{e}^{i 2 (f - \langle \eta_j, \varsigma \rangle t)} \ d t
\ .
\ee 
Since $g_2$ is small with respect to the other frequencies of the system (see section~\ref{sec:orb}), and have for sole effect to slightly offset the frequency $n$, we consider from now on that $n-g_2 \approx n$. We refer to the spin-orbit resonances located in $\dot{\theta} = pn$, i.e. those with $\eta_j=(p,0,0)$, as ``eccentric spin-orbit resonances", and ``coorbital spin-orbit resonances" those located in $\dot{\theta}= pn \pm q \nu$ ($\eta_j=(p,q,0)_{q\neq0}$). 
Replacing equation (\ref{eq:asreqp}) into equation (\ref{eq:thepp}) and denoting $\rho_{\eta_j}\equiv \sigma\hat{\rho}_{\eta_j}$, we get:
\begin{equation}
\label{eq:theppqp} 
\gH=T +\frac{I^2}{2} -  \sum_{j \geq 0 }\ \frac{\rho^2_{\eta_j}}{2} \cos\ (2\theta + 2 \langle \eta_j, \varsigma \rangle t+ \phi_j).
\end{equation}
 $\gH$ can be seen as the Hamiltonian of a quasi-periodically perturbed pendulum, whose forcing frequencies are $n$, $\nu$, $g$ and $g_2$. This is a generalisation of the results by \cite{1965C} and \cite{1966GP}, who showed that there is a whole family of eccentric spin-orbit resonances centred at $\dot{\theta}=pn$ in the eccentric Keplerian case. Here we have a much larger family of spin-orbit resonances centred at $\dot{\theta}=\langle \eta_j, \varsigma \rangle$ with half-width $\rho_{\eta_j}$. 
Since we generally have $n \gg \nu \gg g \gg g_2$, as long as $\sigma \ll n$, i.e. as long as the resonances do not overlap, it is possible to study the rotational dynamics in the vicinity of a particular eccentric spin-orbit resonance ($\dot{\theta}=pn$), because it is not perturbed by the other eccentric spin-orbit resonances.  
Following \citet{1966GP}, one can get an approximated integrable equation for the rotation in the vicinity of $\dot{\theta}=pn$ by studying the angle $\gamma_p$, defined as:
\be
\gamma_p \equiv \theta - pnt,
\ee
where $p$ specifies the eccentric spin-orbit resonance that we are studying. 
For example, $p=1$ in the case of the synchronous resonance like the case of the Moon, 
and $p=3/2$ in the case of the $3$:$2$ resonance of Mercury.

In order to adopt an autonomous Hamiltonian formulation which takes into account the different time-scales, we denote $\lambda=nt$, $\tau=\nu t$, $\upsilon=gt$ and $\upsilon_2=g_2t$. 
We recall that $n$ is the mean mean motion of the system, $\nu=\gO(\sqrt{\mu})$ is the fundamental frequency of the libration angle $\zeta$, and $g=\gO(\mu)$ and $g_2=\gO(\mu)$ are the frequencies of libration and precession of the pericenter, respectively. 
In our study we consider only slightly eccentric orbits, and $\mu \ll1$, we can thus neglect the terms of order $\mu$, $e^2$ and $\sqrt{\mu}e$ or above in the perturbation. 
The solution for $\zeta$ in equation (\ref{eq:sol_orb_zeta}) can be rewritten as:
\be
\zeta(t)=\zeta(\tau/\nu)=\hat{\zeta}(\tau).
\ee  
In the case of co-orbital bodies, at first order in eccentricity, injecting (\ref{eq:lam_sol}) into equations (\ref{eq:polar}) we have:
\be
f=f_0+nt+\delta\hat{\zeta}(\tau)+2e(\upsilon)\sin u,
\ee
with $u=\lambda-\varpi(\upsilon,\upsilon_2)+\delta\hat{\zeta}(\tau)$, $\delta = \frac{m_2}{m_1+m_2}$, and
\be
\left(\frac{\bar{a}}{r}\right)^3=1+h(\lambda,\tau,\upsilon)+\gO(\mu,e\sqrt{\mu}),
\ee
where
\be
h(\lambda,\tau,\upsilon)=2\frac{\nu}{n}\delta\hat{\zeta}'(\tau)-3e(\upsilon) \cos u.
\ee

Using the conjugated variables $(\gamma_p,\Gamma_p)=(\theta-pnt,\dot{\theta}-pn)$, $(\lambda,\Lambda)$, $(\tau, T)$, $(\upsilon, \Upsilon)$ and $(\upsilon_2, \Upsilon_2)$ the Hamiltonian (\ref{eq:thepp}) becomes:
\be
\gH=\gH_0+\gH_1+\gO(\mu,\sqrt{\mu}e,e^2),
\label{eq:Hgen}
\ee
with
\be
\gH_0=n\Lambda+\nu T+g \Upsilon+ g_2 \Upsilon_2 +\frac{\Gamma_p^2}{2},
\ee
and
\be
\gH_1=-\frac{\sigma^2}{4}(1+h(\lambda,\tau,\upsilon)) \cos 2(\gamma_p+(p-1)\lambda-\varpi(\upsilon,\upsilon_2)-\delta\hat{\zeta}(\tau)-2e(\upsilon) \sin u).
\label{eq:H1gen}
\ee

The Hamiltonian $\gH$ describes the motion of a rotating asymmetric body in a perturbed Keplerian orbit, at order $1$ in $e$.
As in equation (\ref{eq:theppqp}), we can develop $\gH$ in a Fourier series, which gives: 
\begin{equation}
\gH=\gH_0-\sum_{j\geq0}\frac{\rho^2_{\eta_j}}{4} \cos(2\gamma_p +2 \langle \eta_j, \varsigma \rangle t +\phi_j)+\gO(\mu,\sqrt{\mu}e,e^2).
\end{equation}

\subsection{Dynamical regimes}
\label{sec:DR}
 Following \cite{1979CB}, the dynamics of the rotating body depends on the width of the resonant island $\rho_{\eta_j}$ and on the distance between the center of the resonant islands. 
We define the distance between the center of two resonant islands as:
\be
 \epsilon^k_j=|\langle \eta_j-\eta_k, \varsigma \rangle|.
\ee
Chirikov's resonance overlap criterion states that when the sum of the two unperturbed half-widths is commensurate to the separation of the resonance center ($\epsilon^k_j\approx \rho_{\eta_j}+\rho_{\eta_k}$), large-scale chaos ensues (\cite{1979CB}; \cite{1999ssd}) as it is the case for the rotation of Hyperion (\cite{1984W}) in the eccentric Keplerian case.

 When $(\rho_{\eta_j}+\rho_{\eta_k}) \ll \epsilon^k_j$ the islands are isolated and only small areas near the separatrix of the pendulums are chaotic. Most of the trajectories between those island are quasi-periodic. If we consider a weak dissipation, the rotation can be captured in one of those resonances. The wider the resonant island is, the higher are the probabilities of being captured in it (\cite{1966GP}). 
 
Finally if $(\rho_{\eta_j}+\rho_{\eta_k}) \gg \epsilon^k_j$, resonant islands totally overlap and we get the dynamics of a modulated pendulum \citep[see][page 222]{2002M}. This situation is likely to occur when the Keplerian motion is perturbed by secular phenomena, for example by the precession of the orbits in the planetary case.

In the co-orbital case, the libration is on a semi-secular time scale. Let us consider two resonances separated by $\epsilon^k_j=\nu/2$ (for example $\eta_j=(1,1/2,0)$ and $\eta_k=(1,0,0)$). As we will see in the following sections, $\rho_{\eta_j}/\sigma\leq1$, thus $(\rho_{\eta_j}+\rho_{\eta_k}) < 2\sigma$. 
In general, we have $\sigma/n < 1$ and $\nu/n \leq 0.1$, depending on $\mu$ and $\zeta_0$ \citep{LeRoCo2015}. 
As a consequence, the resonant islands of two resonances separated by $\epsilon^k_j=\nu/2$ can either be isolated, overlapping, or simply reduced to a single modulated resonance, depending on the values of $\rho_{\eta_j}$ and $\nu$. 
Our purpose is thus to estimate the width of the spin-orbit resonances induced by the orbital frequencies $n$, $\nu$ and $g$.

\subsection{Unperturbed Keplerian Orbits}
\label{sec:ecckep}

One can obtain the Hamiltonian for the rotation of an asymmetric body on a single Keplerian orbit by simplifying the Hamiltonian $\gH$ (Eq.~\ref{eq:Hgen}). 
Indeed, by taking the mass of the perturbing body equal to $0$ ($\delta=0$), we obtain
\be
\gH_K= \gH_0 -\frac{\sigma^2}{4}(1-3e \cos \lambda) \cos 2(\gamma_p+(p-1)\lambda-2e \sin \lambda) +\gO(e^2).
\label{eq:Hk}
\ee

In order to study the spin dynamics near a specific eccentric spin-orbit resonance, we can fix $p$. We can thus write $\gH_K$ as:
\be
\begin{split}
\gH_K= \gH_0(\Gamma_p,\Lambda) + \gH_{1,p} (\gamma_p,\lambda) +\gO(e^2),
\end{split}
\label{eq:gamapecc}
\ee
with $\gH_{1,p}$ a periodic function of $\lambda$. We can thus develop $\gH_{1,p}$ as a Fourier series:  
\be
\begin{split}
\gH_{1,p}(\gamma_p,\lambda)=\overline{\gH}_{1,p}(\gamma_p) + \sum_{k\neq0} C_k(\gamma_p) \cos (k\lambda),
\end{split}
\label{eq:gamapecc}
\ee
with $k$ an integer, and
\be
\overline{\gH}_{1,p}=-\frac{\sigma^2}{4}X^{-3,2}_p(e) \cos 2\gamma_p +\gO(e^2).
\ee
$X^{-3,2}_p(e) = \gO(e^{2|p-1|})$ are the Hansen coefficients , given by:
\be
\left( \frac{r}{a} \right)^{k_1} \operatorname{e}^{i k_2 f} = \sum_{2p \in \mathbb{Z}} X^{k_1,k_2}_p \operatorname{e}^{i 2p M}
\label{eq:hansen}
\ee
where $M$ is the mean anomaly of the rotating body \citep{Hansen1855}. As we limited our description to the first order of eccentricity, only three of the first order resonances have no-null coefficients: $X^{-3,2}_{1}(e)=1 + \gO(e^2)$, $X^{-3,2}_{1/2}(e)= -\frac{e}{2} + \gO(e^2)$ and $X^{-3,2}_{3/2}(e)=\frac{7e}{2} + \gO(e^2)$.

For each individual eccentric resonances, we have $|\dot{\gamma}_p| \ll n $. 
We can thus average $\gH_K$ over $\lambda$ to get
\be
\begin{split}
\overline{\gH}_K= \gH_0(\Gamma_p,\Lambda,T) + \overline{\gH}_{1,p}(\gamma_p) +  \gO(e^2).
\end{split}
\label{eq:gamapecc2}
\ee
We can recognize in the expression of $\overline{\gH}_K$ the first terms of the family of the eccentric spin-orbit resonances located in $\dot{\theta}=pn$ (\cite{1966GP}). For these resonances, we have $\rho_{(p,0,0)}=\sigma \sqrt{|X^{-3,2}_p(e)|}$.

\section{Coorbital bodies with constant eccentricity}
\label{sec:CoC}

\label{sec:DA}

We now consider the co-orbital perturbations on a eccentric Keplerian orbit, for which $e$ and $\varpi$ can be considered as constant, either because their variations are too slow ($g= \mathcal{O}(\mu)$), or because the amplitude of these variations can be neglected. 
We thus consider only the sets $\eta_j=(p,q,0)=(p,q)$. Let us express the equation of rotation (\ref{eq:thepp}) as a function of $\gamma_p$ and $\zeta$. Taking $\varpi =0$, $\gH_1$ (equation (\ref{eq:H1gen})) becomes:  
  \be
\gH_1=-\frac{\sigma^2}{4}(1+h(\lambda,\tau)) \cos 2(\gamma_p+(p-1)\lambda-\delta\hat{\zeta}(\tau)-2e \sin u),
\label{eq:H1coo}
\ee
where
\be
h(\lambda,\tau)=  \frac{2 \nu \delta \hat{\zeta}'(\tau)}{n} - 3 e\cos(\lambda+ \delta \hat{\zeta}(\tau)).
\label{eq:asr3}  
\ee
Similarly to the Keplerian case, at order one in $e$ we can chose $p \in \{1/2,1,3/2\}$ and average $\gH_1$ over $\lambda$, obtaining:

\be
\overline{\gH}_{1,p}= - \frac{\sigma^2}{4}  [ X^{-3,2}_p(e)+ H_c(p)] \cos(2\gamma_p - 2p\hat{\zeta}(\tau)),
\label{eq:resoeq}
\ee
where the Hansen coefficients $X^{-3,2}_p(e)$ are identical to the Keplerian case, $H_c(1)=-\frac{2\delta \nu \hat{\zeta}'}{n}$ and $H_c(1/2)=H_c(3/2)=0$.

Equation (\ref{eq:sol_orb_zeta}) is not integrable, so in general we cannot obtain an analytical expression of the $\rho_{\eta_j}$ in the co-orbital case. Nevertheless, when $\zeta_0$ is near to the Lagrangian equilibrium $\zeta_L$, we can find a simple approximate expression for $\zeta$, and express the $\rho_{\eta_j}$ as functions of the initial parameters. 

\subsection{Near the Lagrangian Circular Equilibrium}
\label{sec:LM}
We use a development of order $2$ to understand the behaviour of the system for $\zeta_0$ near $\zeta_L$, $i.e.$, in the vicinity the Lagrangian circular equilibrium point (we recall that $\zeta_0$ is the minimal value reached by $\zeta(t) =\lambda_1(t) - \lambda_2(t)$ during a libration period, and $\zeta_L = \pi/3$ for the $l_4$ equilibrium and $5\pi/3$ for the $l_5$ one).
We define $z=\zeta_0 - \zeta_L$, the departure from the Lagrangian equilibrium at $t=0$ and $\frac{d \hat{\zeta}}{d \tau} = 0$. A quadratic approximation of the solution of the equation (\ref{eq:sol_orb_zeta}) is:
\be
\label{eq:zetaap}
\hat{\zeta}(\tau)= \zeta_L+\frac{3\sqrt{3}}{8}z^2 + (1-\frac{\sqrt{3}}{4}z)z \cos ( \tau ) -\frac{\sqrt{3}}{8}z^2 \cos (2\tau) +\gO(z^3).
\ee
\vspace{10pt}
The constant term $\zeta_L+\frac{3\sqrt{3}}{8}z^2$ can be absorbed by a redefinition of $\lambda$ and $\gamma_p$, we thus neglect it from now on. At second order in $z$, the averaged part of the Hamiltonian $\overline{\gH}_{1,p}$, equation (\ref{eq:resoeq}), becomes, for $p=1$:
\be
\begin{split}
\overline{\gH}_{1,p=1}= - \frac{\sigma^2}{4}\left[ (1-z^2\delta^2) \cos(2\gamma_1) +z\delta(1-\frac{\sqrt{3}}{4}z -\frac{\nu}{n})\cos(2\gamma_1+\tau-\frac{\pi}{2}) \right.\\
\left.+z\delta(1-\frac{\sqrt{3}}{4}z+\frac{\nu}{n})\cos(2\gamma_1-\tau-\frac{\pi}{2}) \right.\\
\left.+\frac{z^2\delta}{2}\sqrt{\delta^2+\frac{3}{16}}\cos(2\gamma_1+2\tau+\phi) \right.\\
\left.+\frac{z^2\delta}{2}\sqrt{\delta^2+\frac{3}{16}}\cos(2\gamma_1-2\tau+\phi) \right]+ \gO (e^2,ez^2,z\mu ),
\end{split}
\label{eq:gamap}
\ee
\vspace{10pt}
with $\phi$ a phase depending on $z$ and $\delta=m_2/(m_1+m_2)$, and for $p=3/2$ or $1/2$:
\be
\begin{split}
\overline{\gH}_{1,p}= - \frac{\sigma^2}{4} X^{-3,2}_p(e) \left[ \cos(2\gamma_p) +p z\delta\cos(2\gamma_p + \tau) \right.\\
\left. +p z\delta\cos(2\gamma_p - \tau) \right]+ \gO(e^2,ez^2,z\mu ).
\end{split}
\label{eq:gamap}
\ee

First, let us consider these results at first order in $e$ and $z$. 
If $e=0$, we get the expression in the quasi-circular case, as developed by \cite{2013CR}. In that paper, the authors studied the overlap between the synchronous resonant island $\eta_j=(1,0)$ and the island centred at $\eta_j = (1, \pm 1/2)$. Here we generalise the study to the vicinity of the eccentric spin-orbit resonances $\eta_j=(1/2,0)$ and  $\eta_j=(3/2,0)$.

 In the unperturbed eccentric case, we recalled in the previous section that there were eccentric spin-orbit resonances centred at $\eta_j=(p,0)_{p \in\{1/2,1,3/2\}}$. The libration of amplitude $z$ around the Lagrangian equilibrium splits each of these eccentric resonances into three resonant islands as well. Equation (\ref{eq:gamap}) shows that we can approximate the phase portrait in the vicinity of each of the unperturbed eccentric spin-orbit resonances ($1$:$1$, $3$:$2$ or $1$:$2$) by an island of half-width $\rho_{(p,0)}=\sqrt{X^{-3,2}_p(e)}\sigma$ centred at $\eta_j=(p,0)$ and two island of half-width $\rho_{(p,\pm1/2)}=\sqrt{pz\delta X^{-3,2}_p(e)}\sigma$ located at $\eta_j=(p,\pm 1/2)$ (see Figure \ref{fig:fm32a}). The contribution of $\hat{\zeta}'$ in the expression of $h$ (equation (\ref{eq:asr3})) is of order $\nu/n=\gO(\sqrt{\mu})$, and we will neglect it from now on.

We can estimate for which values of the axial asymmetry of the rotation body $\sigma$ the resonant islands in $\eta_j=(p,0)$ and $\eta_k=(p, - 1/2)$ begin to overlap. We recall that near the Lagrangian equilibrium, $\nu=n\sqrt{27\mu}/2$. Therefore, for these resonances $\epsilon^k_j=n\sqrt{27\mu}/4$ and $(\rho_{\eta_j}+\rho_{\eta_k})=(1+\sqrt{z\delta})\sigma\sqrt{X^{-3,2}_p(e)}$. 
By using the definition of $\sigma$ (equation (\ref{eq:sig1})), one can see that these resonant islands overlap when
\be
\frac{B-A}{C} \approx \frac{9}{16\sqrt{X^{-3,2}_p(e)}}(1-2\sqrt{z\delta})\mu \ .
\ee
The interaction between two islands of first order in $z$ is also theoretically possible, for example the islands $\eta_j=(1,1/2)$ and $\eta_k=(3/2,-1/2)$ (see Figure \ref{fig:fmrBM}). However, in order to get those two islands near to each other, we need to have $\nu\approx n/2$, which is near the stability limit $\nu=n/\sqrt{2}$ given by \cite{1843GG}. 
Moreover, in order to interact, these two islands also need to present large libration widths, but a simultaneously large amplitude $z$ and a high frequency $\nu$ is not possible for stability reasons \citep{LeRoCo2015}.

Due to the truncation of the terms of order $ez^2$, only the synchronous resonance is modified by the terms in $z^2$ in this model. 
For large libration amplitudes $z$, co-orbital resonant islands of width proportional to $z^2$ appear in $\dot{\theta}=n\pm\nu$, while the width of the synchronous island decreases (see Fig \ref{fig:Kr} and \ref{fig:fm32a}). 
In next section we will see that the trends of the quadratic model are confirmed for larger libration amplitudes, and a similar behaviour occurs in the vicinity of all eccentric spin-orbit resonances.

\subsection{Tadpole and Horseshoe Configurations}
\label{sec:SAS}

\begin{figure}[h!]
\begin{center}
\includegraphics[width=5.5cm,height=5.6cm]{\figpath/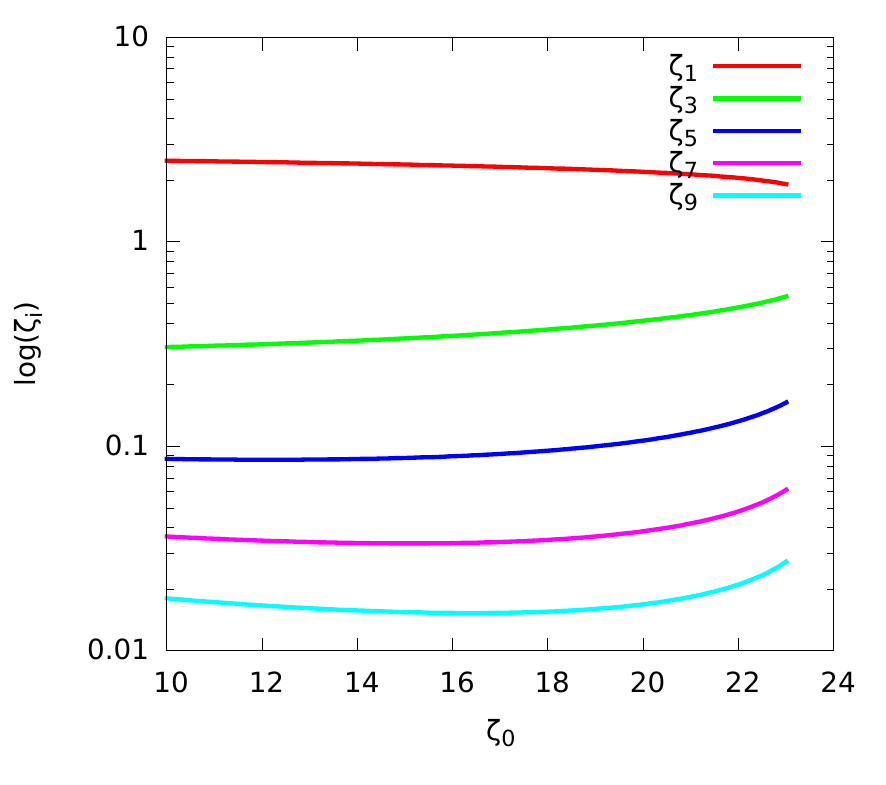} 
\includegraphics[width=5.5cm,height=5.6cm]{\figpath/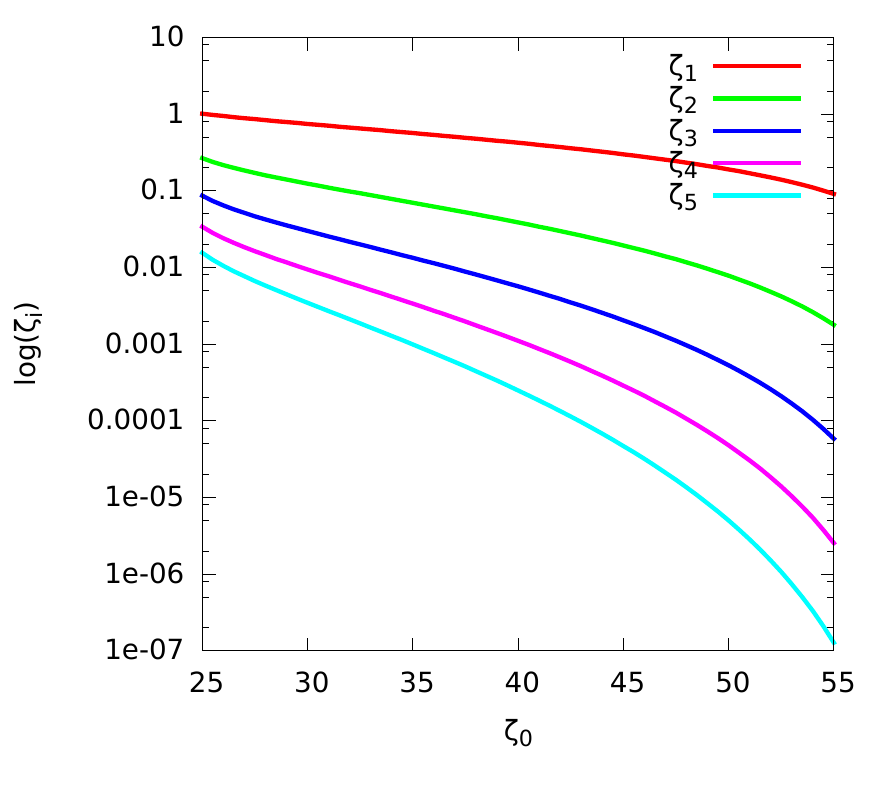} 
\end{center}
\caption{The five dominant coefficients of the Fourier expansion of $\hat{\zeta}$, equation (\ref{eq:deczet}). The $\xi_l$ have been obtained by a numerical integration of equation (\ref{eq:sol_orb_zeta}) followed by a frequency analysis. They depend only on $\zeta_0$. On the left, in the horseshoe case, the symmetry of the orbit imposes that $\xi_{2l}=0$ \citep{RoRaEl2012}. On the right, we represented those coefficients for the tadpole configuration.}
\label{fig:zetai}
\end{figure}
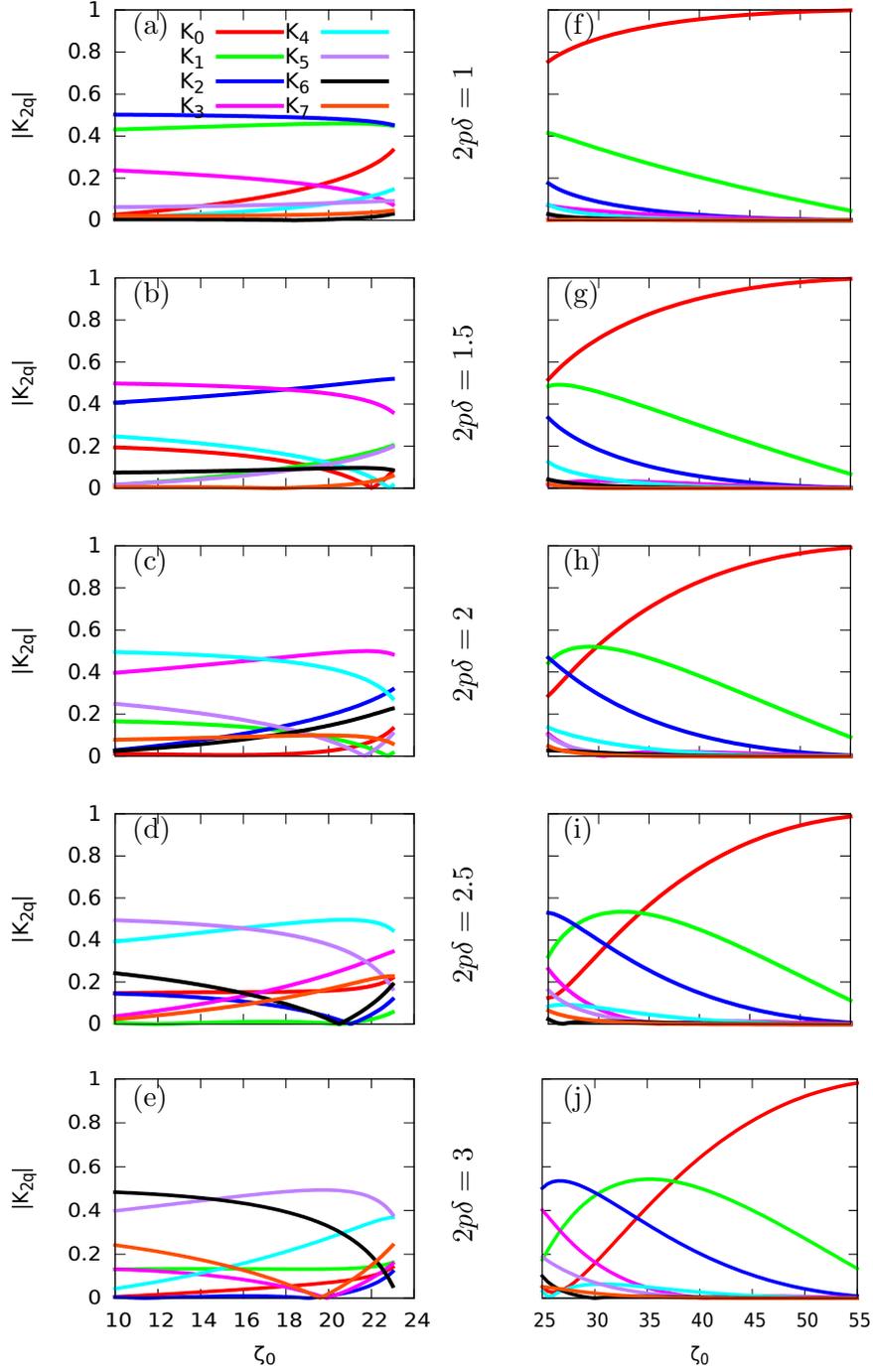
\begin{figure}
\begin{center}
\includegraphics[width=6cm,height=3.5cm]{\figpath/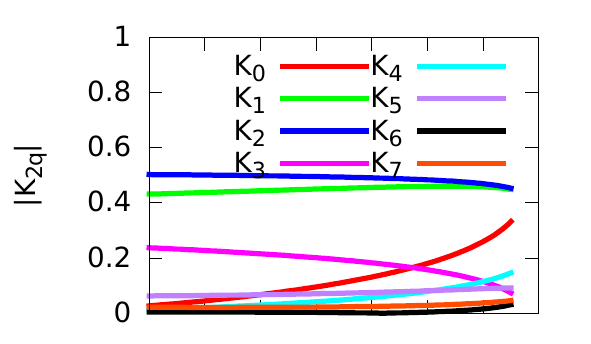} 
\hspace{15pt}
\includegraphics[width=5cm,height=3.5cm]{\figpath/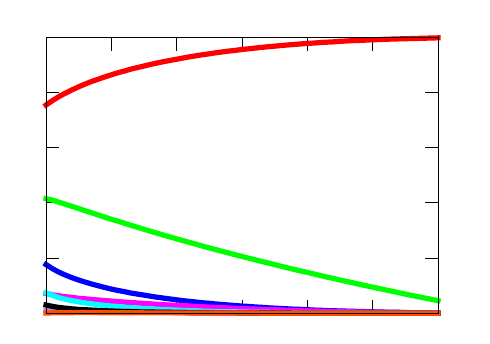}  \setlength{\unitlength}{1cm}
\begin{picture}(.01,.01)
\put(-10.3,2.8){(a)}
\put(-4.6,2.8){(f)}
\end{picture}
\end{center}
\vspace{-0.8cm}
\begin{center}
\includegraphics[width=6cm,height=3.5cm]{\figpath/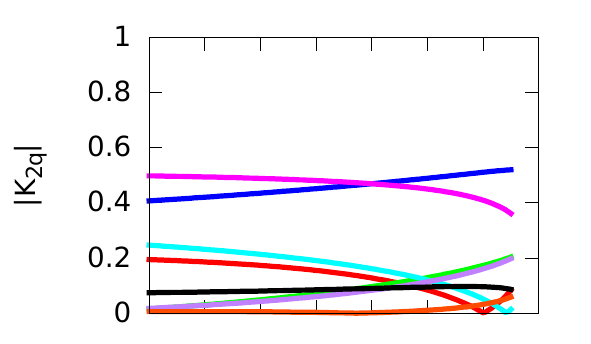} 
\hspace{15pt}
\includegraphics[width=5cm,height=3.5cm]{\figpath/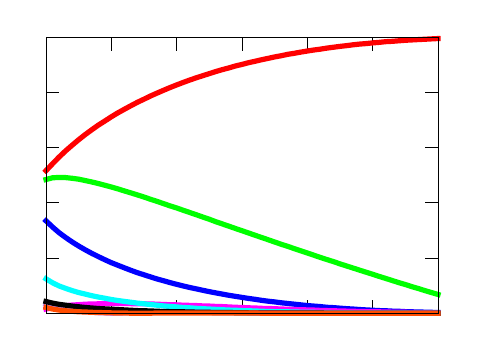} \setlength{\unitlength}{1cm}
\begin{picture}(.01,.01)
\put(-10.3,2.8){(b)}
\put(-4.6,2.8){(g)}
\end{picture}
\end{center}
\vspace{-0.8cm}
\begin{center}
\includegraphics[width=6cm,height=3.5cm]{\figpath/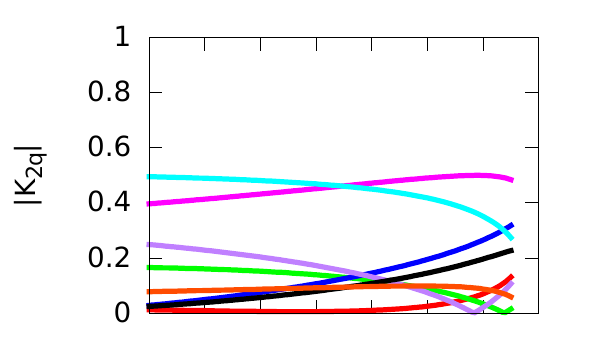} 
\hspace{15pt}
\includegraphics[width=5cm,height=3.5cm]{\figpath/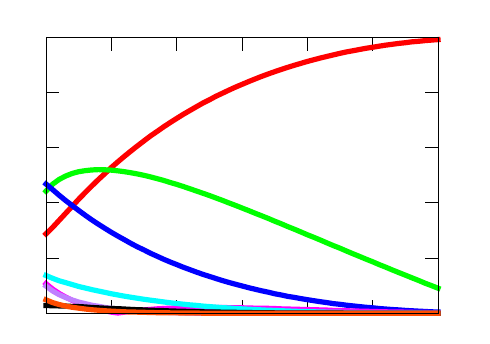} \setlength{\unitlength}{1cm}
\begin{picture}(.01,.01)
\put(-10.3,2.8){(c)}
\put(-4.6,2.8){(h)}
\end{picture}
\end{center}
\vspace{-0.8cm}
\begin{center}
\includegraphics[width=6cm,height=3.5cm]{\figpath/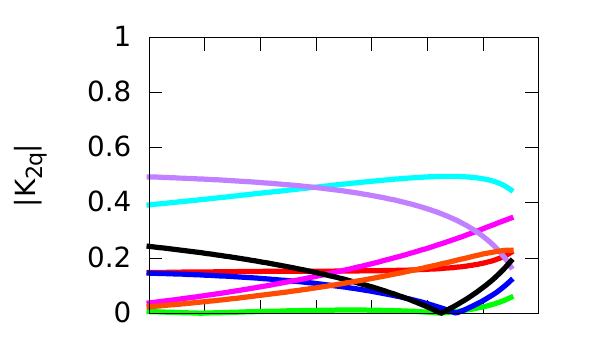} 
\hspace{15pt}
\includegraphics[width=5cm,height=3.5cm]{\figpath/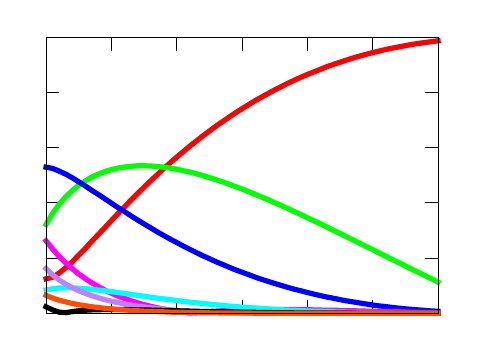} \setlength{\unitlength}{1cm}
\begin{picture}(.01,.01)
\put(-10.3,2.8){(d)}
\put(-4.6,2.8){(i)}
\end{picture}
\end{center}
\vspace{-0.8cm}
\begin{center}
\includegraphics[width=6cm,height=4.5cm]{\figpath/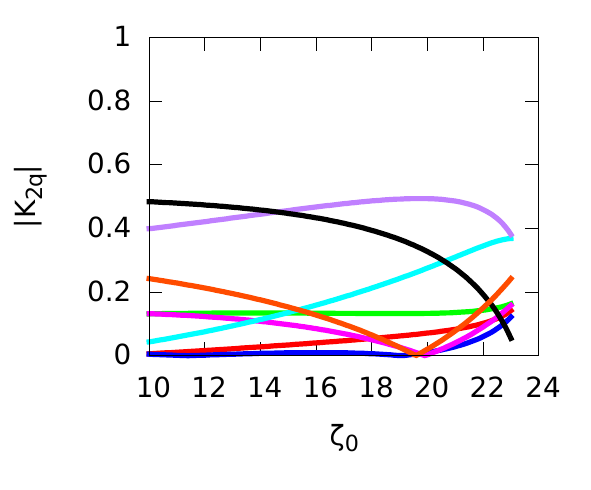} 
\hspace{15pt}
\includegraphics[width=5cm,height=4.5cm]{\figpath/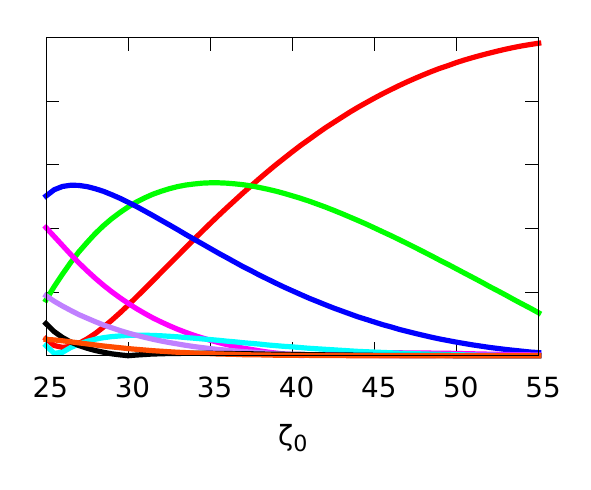} \setlength{\unitlength}{1cm}
\begin{picture}(.01,.01)
\put(-10.3,3.8){(e)}
\put(-4.6,3.8){(j)}
\end{picture}
\end{center}
\vspace{-.5cm}
 \setlength{\unitlength}{1cm}
\begin{picture}(0.001,0.001)
\hspace{175pt} 
\rotatebox{90}{ \hspace{60pt} $2p \delta = 3$
\hspace{55pt} $2p \delta = 2.5$
\hspace{55pt} $2p \delta = 2$
\hspace{55pt} $2p \delta = 1.5$
\hspace{55pt} $2p \delta = 1$
}
\end{picture} 
\vspace{-0.5cm}
\caption{The seven firsts $|K_{2q}|_{2q \in \mathbb{Z}}$ coefficients with respect to $\zeta_0$ for the horseshoe on the left and for the tadpole orbits orbits on the right. For (a) \& (f) we have $2p\delta = 1$, for (b) \& (g) $2p\delta=3/2$, for (c) \& (h) $2p\delta=2$, for (d) \& (i) $2p\delta=5/2$ and for (e) \& (j) $ 2p\delta=3$. see the text for more details.}
\label{fig:Kr}
\end{figure}

We now describe the spin dynamics of the coorbital in the whole tadpole and horseshoe domains. To do so, we write the averaged part of the Hamiltonian $\overline{\gH}_p$ (\ref{eq:resoeq}) under the form:
\begin{equation}
\overline{\gH}_p=\gH_0-\sum_{q\in\mathbb{Z}}\frac{\rho^2_{(p,q)}}{4} \cos(2\gamma_p + q\tau+\phi),
\label{eq:gHp}
\end{equation}

The explicit expression of this Hamiltonian given in section \ref{sec:LM} is no more valid for $\zeta_0 - \zeta_L \gg 1$. In order to have a valid expression everywhere in the tadpole and horseshoe domain, we consider the fourier expansion of the solution $\hat{\zeta}(\tau)$  of equation (\ref{eq:resoeq}). $\hat{\zeta}$ being a real function of period $2\pi$, we can write:
\be
 \hat{\zeta} =  \xi_0 + \sum_{l \geq 1}\ \xi_l \sin(l \tau +\varphi_l) 
 \label{eq:deczet}
\ee
The coefficients $\xi_l$ can be obtained numerically. In Figure~\ref{fig:zetai} we show the evolution of the lower order $\xi_l$ with respect to $\zeta_0$. 
Replacing equation (\ref{eq:deczet}) into (\ref{eq:resoeq}) and neglecting the terms of order $\sqrt{\mu}$ (see section \ref{sec:LM}), we can write:
\be
\overline{\gH}_{1,p}= - \frac{\sigma^2}{4}  X^{-3,2}_p(e) \Re( \operatorname{e}^{i2\gamma_p} \operatorname{e}^{-i 2p\delta \hat{\zeta}(\tau)}).
\label{eq:resoeqR}
\ee

Since $\operatorname{e}^{-i 2p\delta\hat{\zeta}(\tau)}$ is a periodic function with period $2\pi$ in $\tau$, we can also expand it  in Fourier series\footnote{Note that since $\xi_l$ depends only on $\zeta_0$, the coefficients $K_q$ depend only on the product $p\delta$ instead of $p$ and $\delta$.}
\begin{equation}
\label{eq:deceqdzet}
 \operatorname{e}^{-i2p\delta\hat{\zeta}(\tau)} = \sum_{2q \in \mathbb{Z}}\ K_{2q}(p\delta ,\zeta_0) \operatorname{e}^{  i(2q \tau  +\Phi^p_{2q})} \ ,
\end{equation}
and thus rewrite equation (\ref{eq:resoeqR}) as
\begin{equation}
\overline{\gH}_{1,p}=-\frac{\sigma^2}{4}X^{-3,2}_p(e)\sum_{2q\in\mathbb{Z}}K_{2q}(p\delta,\zeta_0) \cos(2\gamma_p + {2q}\tau+\phi) \ .
\label{eq:H1bK}
\end{equation}
By determining the $K_{2q}$ coefficients, we can thus directly obtain the $\rho_{\eta_j}$ from
\begin{equation}
\rho_{(p,q)}=\sigma \sqrt{X^{-3,2}_p(e)K_{2q}(p\delta,\zeta_0)} \ .
\label{eq:rhoj}
\end{equation}

The coefficients $K_{2q}$ are a function of the $\xi_l$ appearing in eq. (\ref{eq:deczet}). 
Replacing equation (\ref{eq:deczet}) into $\operatorname{e}^{-i2p\delta\hat{\zeta}(\tau)}$ gives
\begin{equation}
\label{eq:expiqzet}
\operatorname{e}^{-i2p\delta\hat{\zeta}(\tau)} = \operatorname{e}^{-i2p\delta\xi_0}\operatorname{e}^{-i2p\delta \sum_{l \geq 0} \xi_{l} \sin (l \tau + \phi_{l})} \ ,
\end{equation} 
and thus
\begin{equation}
\label{eq:expiqzet2}
\operatorname{e}^{-i2p\delta\hat{\zeta}(\tau)} = \operatorname{e}^{-i2p\delta\xi_0} \prod_{l \geq 1} \operatorname{e}^{-i2p\delta \xi_{l}\sin (l \tau + \phi_{l})} \ .
\end{equation} 
By using the Bessel functions, we can further write:
\begin{equation}
\label{eq:expiqzet3}
\operatorname{e}^{-i2p\delta\hat{\zeta}(\tau)} = \operatorname{e}^{-i2p\delta\xi_0} \prod_{l \geq 1}\sum_{k \in \mathbb{Z}}\ J_k(-2p\delta \xi_{l})\ \operatorname{e}^{ik(l \tau + \phi_{l})}.
\end{equation} 
where $J_k$ is the $k^{th}$ bessel function. Replacing this last expression into the equation (\ref{eq:resoeqR}), we get:
 \begin{equation}
\label{eq:Kr}
 K_{2q}(p\delta,\zeta_0) =\prod_{j=1}^\infty \ J_{k_j}(-2p\delta \xi_{j}(\zeta_0)), \quad \text{with}\ (k_1,k_2,..,k_j,..)\ /\ \sum_{j=1}^{\infty} j k_j = 2q.
\end{equation}  
We finally obtain an expression for the $\rho_{(p,q)}$ through expression (\ref{eq:rhoj}), which depends on the $\xi_l$ (Eq.~\ref{eq:deczet}). Recalling the propriety of the Bessel function $j_{-k}=(-1)^kJ_k$, we have $|K_{-2q}|=|K_{2q}|$.

In Figure~\ref{fig:Kr} we plot several $K_{2q}(p\delta,\zeta_0)$ coefficients as a function of $\zeta_0$ for different values of $p\delta$. The $\xi_l$ coefficients are obtained by integration of equation (\ref{eq:sol_orb_zeta}) followed by a frequency analysis of the solution with a truncation at $l=10$ (see the exponential decreasing of the $\xi_l$ in $l$, Fig.~\ref{fig:zetai}). The fact that the coefficients $K_{2q}$ depend only on $p$ and $\delta$ by the product $p\delta$ yields an interesting result: the relative width of the resonant islands in the vicinity of the $1/2$ resonance in the restricted case ($\delta = m_2/(m_1+m_2) \approx 1$) and in the vicinity of the 1:1 resonance with equal masses ($\delta =1/2$) are identical within our approximations, since in both case $2p\delta = 1$. However, the width $\rho_j$ of these islands differ because the Hansen coefficients $X^{-3,2}_p(e)$ depend on $p$. 
 
The amplitude of the $K_{2q}$ in Fig.~\ref{fig:Kr} indicates which resonant island is dominating in the vicinity of a given spin-orbit resonance. In the tadpole cases (panels $a$ to $e$), we have most of the time a phase portrait similar to the one in the quasi-circular case \citep{2013CR}, or the one near the elliptic Lagrangian equilibrium (section \ref{sec:LM}), i.e., a dominating island at $\eta_j=(p,0)$ and smaller islands at $\eta_j=(p,\pm q)$, whose width decreases as $q$ increases. The phase portrait of such case can be seen on Fig. \ref{fig:fm32a} (left). Still in the tadpole configuration, far from the Lagrangian equilibrium, several resonant island may become of commensurable width. An example of this is shown in the right panel of the Figure \ref{fig:fm32a}, which corresponds to the same case shown in Fig.~\ref{fig:Kr} (b) for $\zeta_0=25^\circ$. For high values of the product $p \delta$ and far from the Lagrangian equilibrium, the width of the island centred at $\eta_j=(p,0)$ can even been small with respect to the island located in $q=1/2$ or $q=1$ (see for example Fig.~\ref{fig:Kr} (e) for $\zeta_0=25^\circ$).
In the horseshoe configuration, the width of the eccentric spin-orbit resonance is always negligible with respect to the dominating islands at its vicinity. The largest islands are usually located in $\eta_j=(p,\pm q)$, with $q$ different from zero. As the product $p\delta$ increase, the islands further away from the eccentric resonance (thus for higher $q$) become dominating.

   As a result, the spin dynamics in the vicinity of $\eta_j=(p,0)$ is very different between tadpole and horseshoe orbits. In the tadpole configuration, the main resonant island is generally near the eccentric resonance, forming a large chaotic area centred at $\eta_j=(p,0)$ in the case of an overlap \citep{2013CR}. In the horseshoe case, the direct vicinity of the resonance $\eta_j=(p,0)$ is generally regular. In case of overlapping, chaotic areas arise from each side of $\eta_j=(p,0)$, at a distance which increases with the product $p\delta$.  

The overlapping of the resonant island can change the phase portrait of the rotation from separated resonant islands to chaotic rotation and even to a simple modulation of the islands of the unperturbed Keplerian case. Now that we have an expression of the width of the resonant islands centred at $\eta_j=(p,q)$, we can identify which parameters can give rise to a chaotic behaviour.
Departing from a given system, there are two ways to increase the overlap of close-in resonances:

1) widen the resonant islands:  by increasing $\delta$; by increasing the asymmetry ($\frac{B-A}{C}$) of the rotating body; by increasing $e$, which widens all the islands located at $\eta_j=(p,q)_{p\neq1}$.

2) bring the resonant islands closer from each other: $\nu$ is reduced when $\mu$ decreases, since $\nu$ is proportional to $\sqrt{\mu}$ (equation \ref{eq:nu}).

Reducing $\zeta_0$ has the double effect to widen the $\eta_j=(p,q)_{q\neq 0}$ island in one hand, and to reduce $\nu$ and thus bring the islands closer, on the other hand.

\subsection{Numerical simulations}

For a global view of the rotational dynamics we can use the Frequency Map Analysis (FMA) to represent the phase space \citep{1993Laskar, 2001Robutel}. 
This method is particularly adapted to our problem because the trajectories of the considered systems are close to quasi-periodic trajectories and their dynamics involve up to three fundamental frequencies ($n$, $\nu$ and $g$), so we cannot use Poincar\'e sections. 

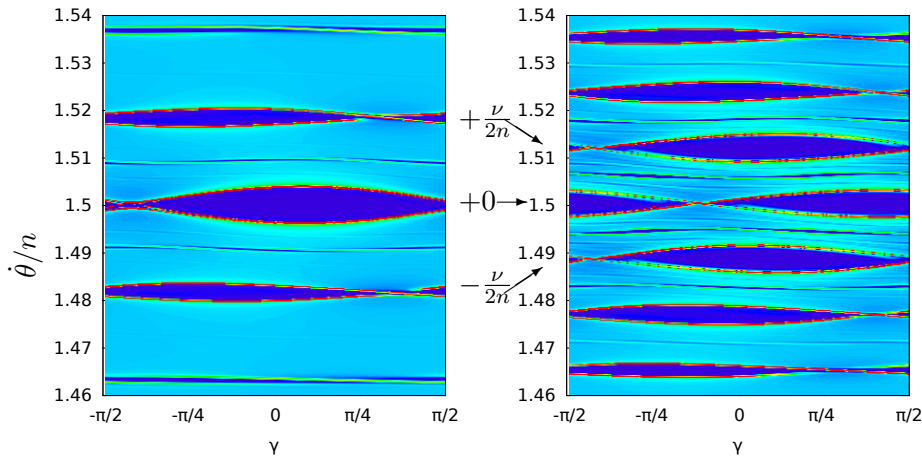
\begin{figure}[h!]
\begin{center}
\rotatebox{90}{\hspace{70pt} $\dot{\theta}/n $}
 \includegraphics[scale=.8]{\figpath/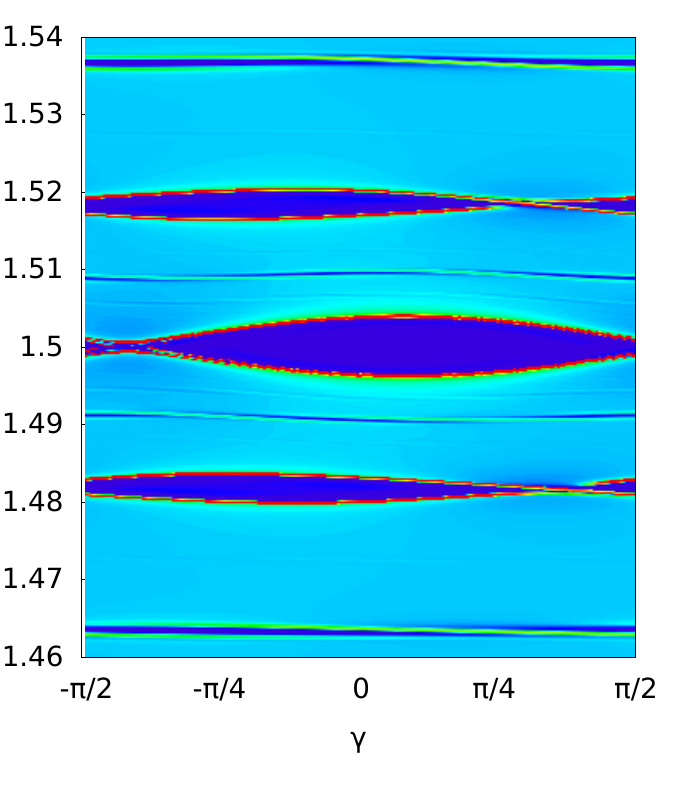}
  \setlength{\unitlength}{1cm}
\begin{picture}(1,0.001)
\put(-0.5,3.5){$+0$}
\put(-0.5,2.4){$-\frac{\nu}{2n}$}
\put(-0.5,4.6){$+\frac{\nu}{2n}$}
\put(0.1,4.7){\vector(3,-2){0.5}}
\put(0.0,3.6){\vector(1,0){0.4}}
\put(0.1,2.4){\vector(3,2){0.5}}

\end{picture} 
\hspace{-1cm}
  \includegraphics[scale=.8]{\figpath/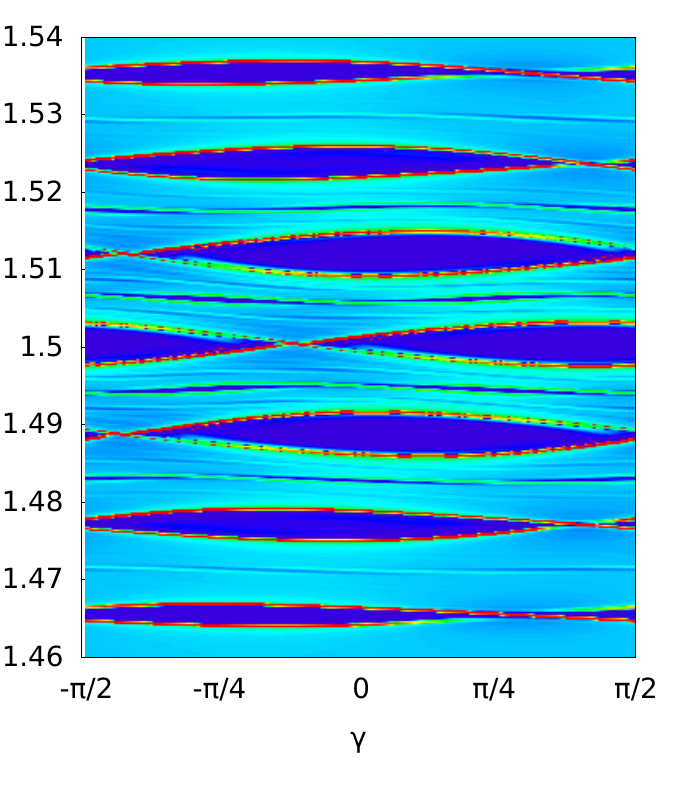}
 \end{center}
 \caption{\label{fig:fm32a} FMA of the phase space ($\dot{\theta}/n,\gamma_p$) for two coorbitals in tadpole configurations. In both cases $m_1=m_2=1\times 10^{-4}\,m_0$, $e_1=e_2=0.1$, and $\frac{B-A}{C}=2\times10^{-5}$. Looking at the vicinity of the $3$:$2$ eccentric spin-obit resonance, we have $2p \delta=1.5$. On the left, we have $\zeta_0=45^\circ$,  $\frac{\nu}{2n}\approx0.018$, the islands are thus located at $\dot{\theta}/n= 1.5 \pm 0.018$. On the right: $\zeta_0=25^\circ$, $\nu/(2n) \approx 0.011$, the islands are thus located at $\dot{\theta}/n= 1.5 \pm 0.011$. $(B-A)/C$ for bodies of the solar system can be found in Appendix \ref{sec:bmasc}, Table 2.}
 \end{figure}

In Figure~\ref{fig:fm32a} we plot one of these maps for the $3$:$2$ resonance.
The vertical axis shows $\dot{\theta}/n$, such that the center of
the $3$:$2$ resonance can be found at $\dot{\theta}/n = 1.5$. 
The horizontal axis is $\gamma=\gamma_1$ modulo $\left[-\frac{\pi}{2},\frac{\pi}{2} \right] $. 
The range of $\gamma$ can be reduced to this interval due to the symmetries of the body. 
The purpose of these simulations is to describe the phase space ($\dot{\theta}/n$,$\gamma$) in order to identify the areas where $\gamma$ librates, circulates or has a chaotic evolution.
Each pixel corresponds to an initial condition in ($\dot{\theta}/n$,$\gamma$) for witch the system is integrated using equation (\ref{eq:thepp}) for the rotational motion together with the equations for the planar three-body problem in heliocentric coordinates. 
The simulations are run over several times the longest time scale considered in the system, generally $\gO(2\pi/\nu)$ or $\gO(2\pi/g)$ (hence $\propto 1/\sqrt{\mu}$ or $\propto1/\mu$ orbital periods, respectively).  For each of those simulations, the main frequency $f$ of $\operatorname{e}^{i \gamma}$ has been computed \citep[see][]{RoRaEl2012}. The color of each pixel represents the first derivative of this frequency with respect to $\dot{\theta}$, i.e., $ d f / d \dot{\theta}$. 
In the circulation region, $f$ evolves almost linearly with respect to $ \dot{\theta}$, thus the value of the derivative is almost constant. Inside the libration island, $f$ is a constant, thus its derivative is equal to zero. Finally, at the separatrix between circulation and libration, and in the chaotic region, $f$ is singular and so is its derivative.

\subsubsection{Tadpole orbits}

We recall that a system is in a tadpole configuration when $\zeta_0 \in (\zeta_s,\zeta_L]$. We saw previously in the quadratic approximation (sections \ref{sec:LM} and \ref{sec:SAS}) that for $\zeta_0$ close to $\zeta_L$, the eccentric spin-orbit resonance ($\eta_j=(p,0)$) is larger than the co-orbital resonances $\eta_j=(p,q)_{q\neq0}$ (see Fig. \ref{fig:fm32a}, left panel). 
Far from the Lagrangian equilibrium, (and especially for high values of $2p\delta$), the width of the resonances $\eta_j=(p,\pm 1/2)$ and $\eta_j=(p,\pm 1)$ (island centred at $\dot{\theta}=pn \pm \nu/2$ or $\dot{\theta}=pn \pm \nu$, respectively) can be as wide as the eccentric islands or even larger (see Fig.~\ref{fig:Kr} and Fig.~\ref{fig:fm32a}, right). Increasing the amplitude of libration gives also rise to higher order resonant island, in-between the previously described resonant island. The appearance of such islands can accelerate the transition to a chaotic phase portrait.

\begin{figure}[]
\includegraphics[width=1\linewidth]{\figpath/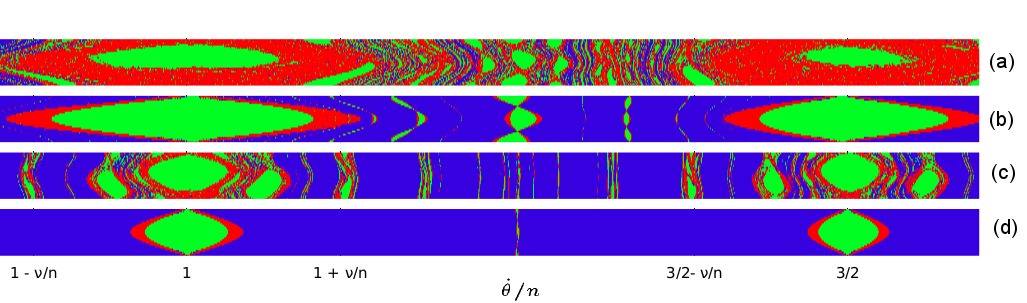}
\caption{\label{fig:fmrBM} Frequency map of focused on the 1:1 to 3:2 resonances area. (a) and (c) represent a co-orbital configuration with $\delta = 1/2$ $\mu=2\times 10^{-3}$, $e=0.15$ and $\zeta_0=45^\circ$. For comparison, (b) and (d) represent the keplerian case ($\delta=0$). $\frac{B-A}{C}=5\times 10^{-4}$ for (c) and (d) and $\frac{B-A}{C}=5\times 10^{-3}$ for (a) and (b). Blue when $\gamma$ circulate, green when it is trapped in a spin-orbit resonance and red when the rotation is chaotic. See the text for more details.}
\end{figure}
For unperturbed Keplerian orbits, chaotic rotation emerges when $e$ and $\sigma$ are large enough, as it is the case for Hyperion \citep{1984W}. 
However, in the co-orbital case, the generalised chaos may have a different origin.
In Figure~\ref{fig:fmrBM} (b) and (d), we show the classic eccentric resonances described in section \ref{sec:ecckep} for unperturbed Keplerian orbits.
In this plot higher order resonances in $e$ are also visible (for example $\dot{\theta}/n=5/4$). 
In Figure~\ref{fig:fmrBM} (a) and (c) we show the previous two orbits while perturbed by a coorbital companion.
We observe that each member of the eccentric spin-orbit family, including the ones of higher order than one in $e$, give rise to co-orbital spin orbit resonances, fulfilling the phase space and potentially overlapping each other. 
Moreover, generalised chaotic rotation can be reached for lower values of $\sigma$ in the co-orbital case than in the unperturbed Keplerian case. 
This is due to a much more populated phase space, ensuring the overlap of all the resonant islands between the 1:1 and the 3:2 resonances.

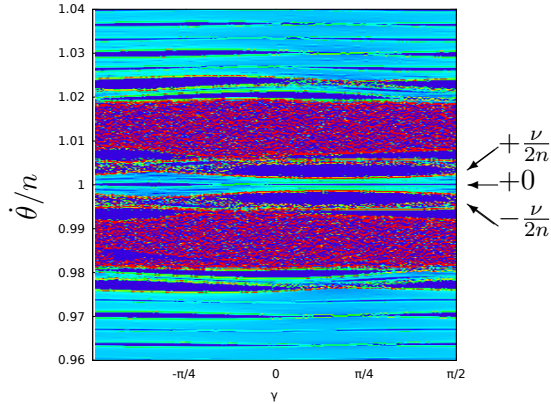
\begin{figure}[h!]
\begin{center}
\rotatebox{90}{\hspace{70pt} $\dot{\theta}/n $}
\includegraphics[scale=0.56]{\figpath/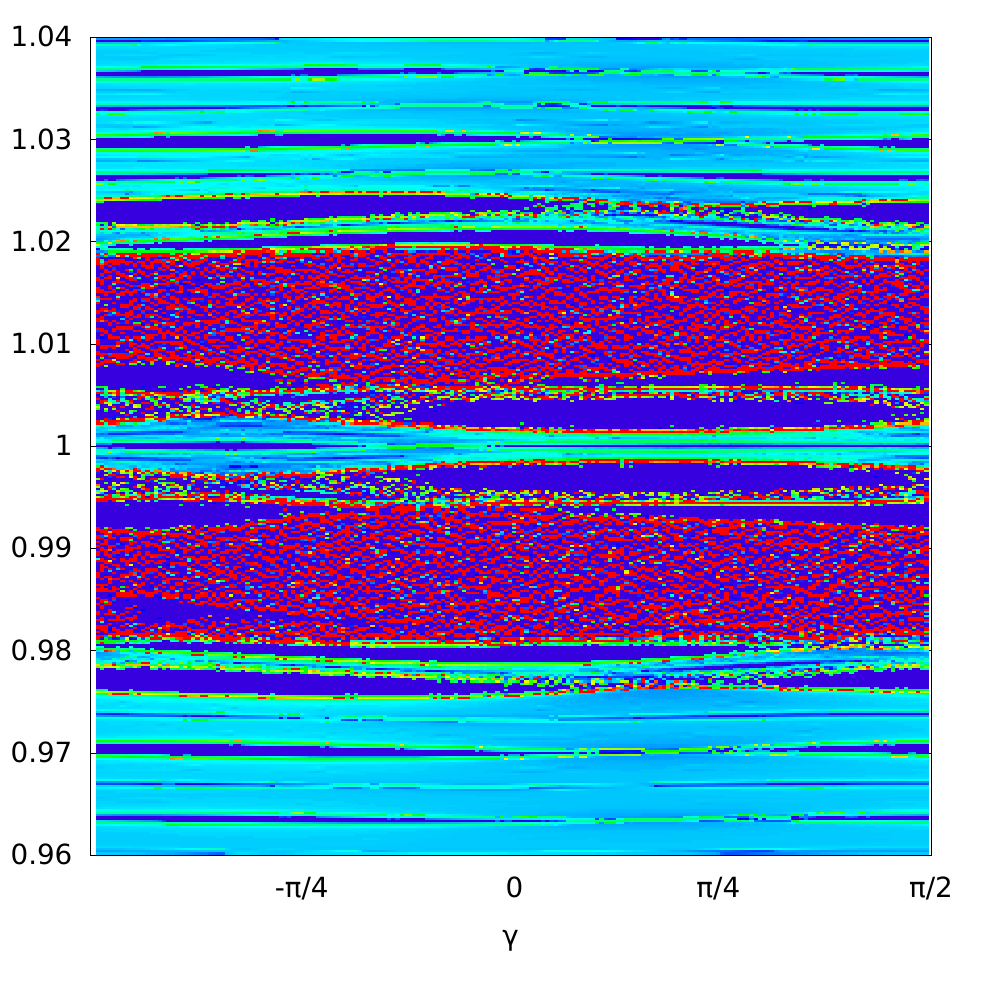} 
  \setlength{\unitlength}{1cm}
\begin{picture}(1,0.001)
\put(0,3){$+0$}
\put(0,2.5){$-\frac{\nu}{2n}$}
\put(0,3.5){$+\frac{\nu}{2n}$}
\put(0,3.05){\vector(-1,0){0.4}}
\put(0,2.5){\vector(-4,3){0.4}}
\put(0,3.55){\vector(-4,-3){0.4}}
\end{picture} 
\end{center}
\caption{\label{fig:fm32b} FMA of the phase space ($\dot{\theta}/n,\gamma_p$) for two coorbitals on horseshoe configuration. $e_1=e_2=0.1$, $\frac{B-A}{C}=1\times 10^{-5}$. $\zeta_0=12^\circ$, $m_1=1\times 10^{-7}m_0$ and $m_2=1\times 10^{-5}m_0$. We focus here on the synchronous resonance with $2p\delta= 2$. $\nu/(2n) \approx 0.0033 $, the islands are thus located at $\dot{\theta}/n= 1 \pm k \times 0.0033$, with $k$ an integer.}
\end{figure}

\begin{figure}[]
\begin{center}
\rotatebox{90}{\hspace{70pt} $\dot{\theta}/n $}
\includegraphics[scale=0.50]{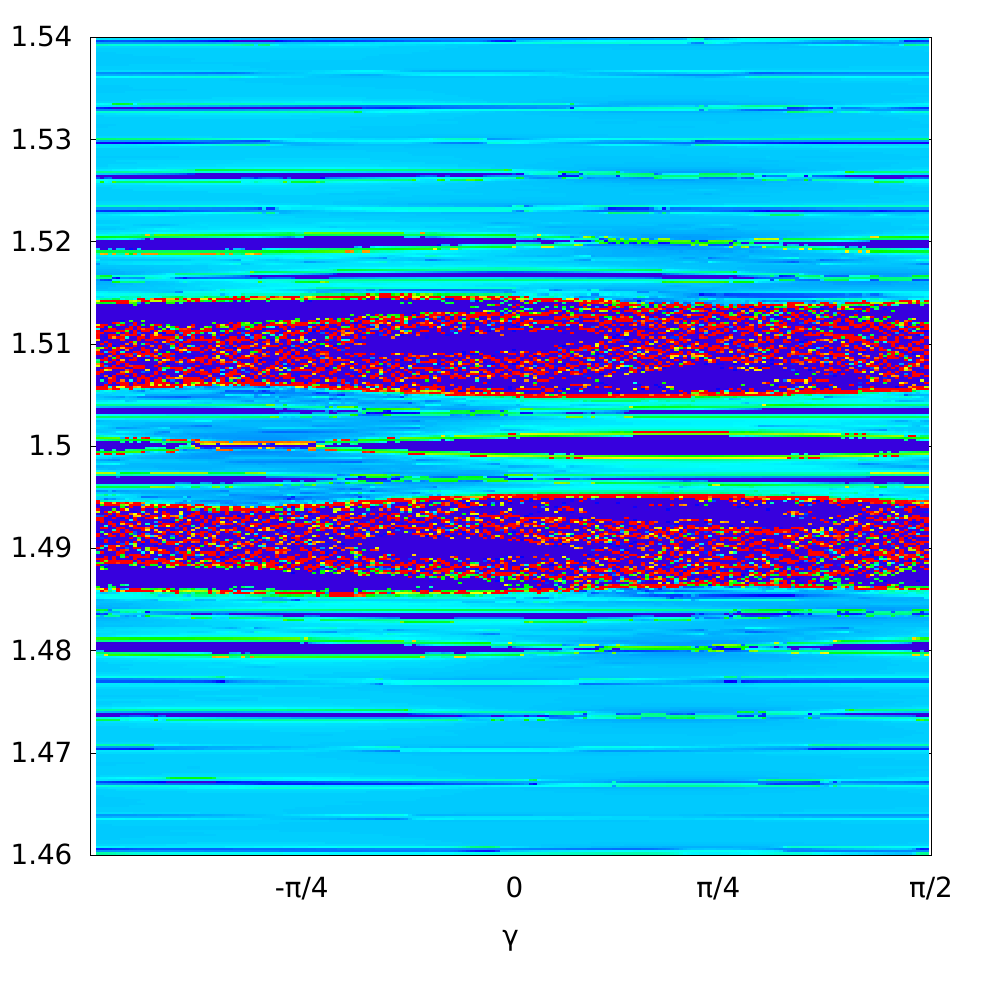} \includegraphics[scale=0.50]{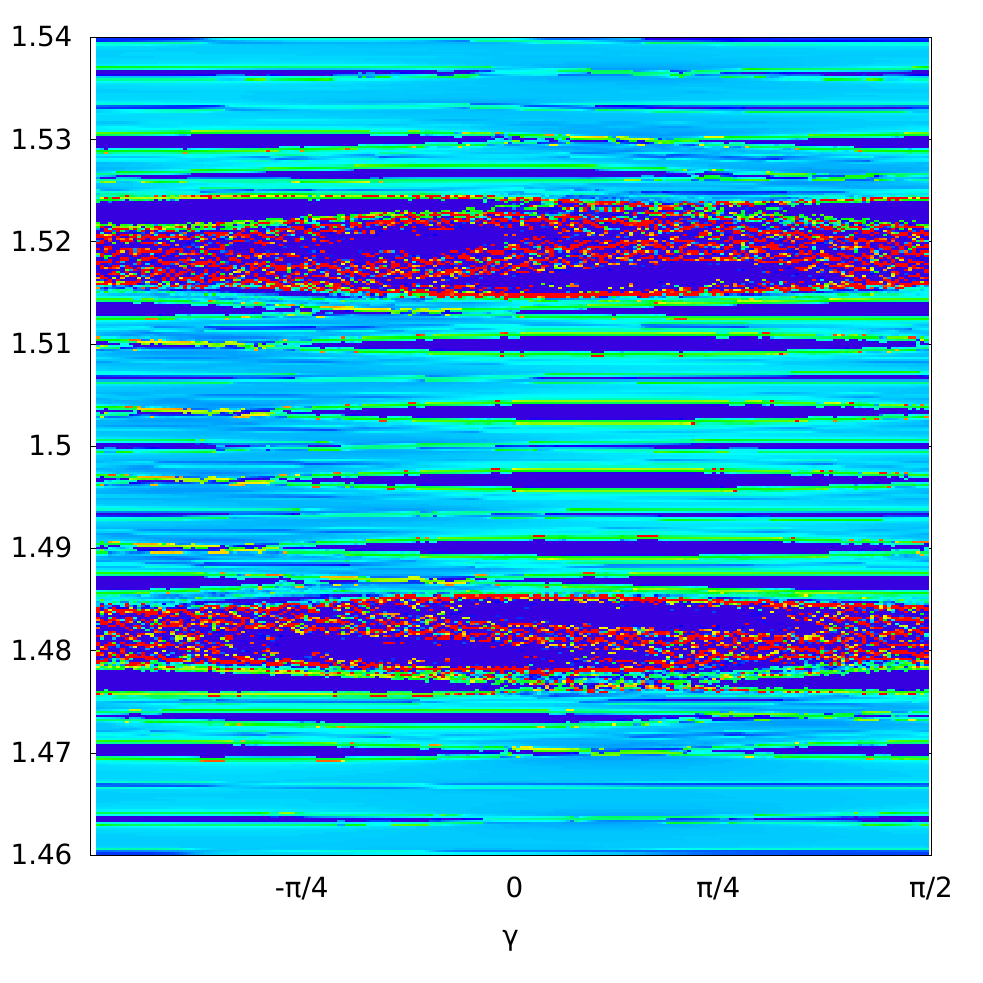} 
  \setlength{\unitlength}{0.892cm}
\begin{picture}(1,0.001)
\put(0,3){$+0$}
\put(0,2.5){$-\frac{\nu}{2n}$}
\put(0,3.5){$+\frac{\nu}{2n}$}
\put(0,3.05){\vector(-1,0){0.4}}
\put(0,2.5){\vector(-4,3){0.4}}
\put(0,3.55){\vector(-4,-3){0.4}}
\end{picture} 
\end{center}
\caption{\label{fig:fm32c} FMA of the phase space ($\dot{\theta}/n,\gamma_p$) in the neighbourhood of the 3:2 spin-orbit resonance for two coorbitals on horseshoe configuration. $e_1=e_2=0.1$, $\frac{B-A}{C}=1\times 10^{-5}$. $\zeta_0=12^\circ$ in both cases. On the left we have $m_1=m_2=5\times 10^{-6}m_0$ ($2p\delta = 3/2$). On the right we have $m_1=1\times 10^{-7}m_0$, $m_2=1\times 10^{-5}m_0$ ($2p\delta = 3$). $\nu/(2n) \approx 0.0033 $, the islands are thus located at $\dot{\theta}/n= 1.5 \pm k \times 0.0033$, with $k$ an integer. See the text for more details.}
\end{figure}

\subsubsection{Horseshoe orbits}

For horseshoe orbits ($\zeta_0 \in (0,\zeta_s)$), the phase portrait is different from tadpole orbits. 
In Figure~\ref{fig:fm32b} we look at the vicinity of the synchronous resonance $\eta_j=(1,0)$, in a case where $m_1 \ll m_2$. As explained in section \ref{sec:SAS}, in the horseshoe configuration the main resonant island is not centred at the synchronous resonance, but on each side, symmetrically with respect to $\eta_j=(1,0)$. In this case, we have $2p\delta=2$ and $\zeta_0=12^\circ$, so the main resonances are for $2q=3$ and $2q=4$ (see Fig.\,\ref{fig:Kr}\,h), which are centred at $\eta_j=(1,\pm 3/2)$ and $\eta_j=(1,\pm 2)$. In Figure~\ref{fig:fm32b} these resonances overlap, creating a chaotic area that encompass nearby resonances.   

As for the synchronous resonance, in the vicinity of an eccentric spin-orbit resonance ($\eta_j=(p,0)_{p \neq 0}$), the main resonant island is not the eccentric resonance itself, since the order $q$ of the dominating harmonics depends mainly on the product $2p\delta$. 
In Figure \ref{fig:fm32c} we look at the vicinity of the 3:2 eccentric spin-orbit resonance for two different values of $\delta$. On the left panel, the two coorbital have the same mass, while on the right, we are in the restricted case $m_1 \ll m_2$. We thus have $2p\delta=3/2$ on the left and $2p\delta=3$ on the right. Accordingly to the graphs (g) and (j) of Fig.~\ref{fig:Kr}, the largest resonant islands are located at $\eta_j=(3/2,\pm 1)$ and $\eta_j=(3/2,\pm 3/2)$ for the equal masses and $\eta_j=(3/2,\pm 5/2)$ and $\eta_j=(3/2,\pm 3)$ in the restricted case. It is indeed what we observe on Figure \ref{fig:fm32c}. In both cases, the main resonant islands overlap, creating two chaotic area symmetric with respect to the eccentric spin-orbit resonance $\eta_j=(3/2,0)$.

Overall, In the horseshoe domain, the eccentric spin-orbit resonances are of negligible width with respect to some of the co-orbital resonant islands under and above them. In addition, since horseshoe co-orbitals are stable only for $\mu < 2 \times 10^{-4}$, the resonant island are generally closer to each other than in the tadpole configuration ($\nu =\gO (\sqrt{\mu})$). That is why, in the three examples we have chosen, the main resonant islands always overlap.

\subsubsection{Quasi satellite}

\begin{figure}[h!]
\begin{center}
\rotatebox{90}{\hspace{35pt} $\dot{\theta}/n $}
 \includegraphics[scale=0.75]{\figpath/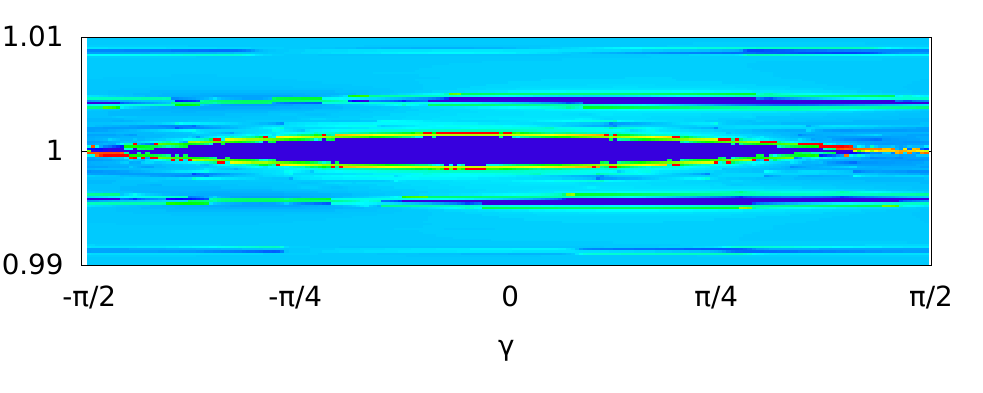}  
\end{center}
\caption{\label{fig:r11qs} FMA of the phase space ($\dot{\theta}/n,\gamma_p$) for two coorbitals in quasi-satellite configuration, near the synchronous resonance. $m_1=m_2=1\times 10^{-6}m_0$, $e_1=e_2=0.2$, $\frac{B-A}{C}=1\times 10^{-6}$, $\varpi_1-\varpi_2=180^\circ$ and $\zeta_0=15^\circ$. See the text for more details.}
\end{figure}

Although our model does not apply to the quasi-satellite case, they have a similar dynamics near the main eccentric spin-orbit resonances. i.e. resonant islands located in $\dot{\theta}= n \pm \nu/2$. In Figure \ref{fig:r11qs} we show the vicinity of the 1:1 resonance for a quasi-satellite case. In this case $\nu/(2 n) =4.3\times 10^{-3}$, thus the islands are located in $\dot{\theta}/n = 1 \pm \nu/(4\pi) = 1 \pm 0.0043$.

\section{Coorbital bodies with variable eccentricity}
\label{sec:evar}

Until now, we have been considering that $e$ and $\varpi$ are constant because either their variations are too small, either they are slow with respect to the considered time-scale. In general, the temporal evolution of $e$ and $\varpi$ depends on $g$ and $g_2$. In the vicinity of the Lagrangian equilibrium, i.e., when the variations of $e$ are small, we can make the following approximation (Eq.~\ref{eq:ew}):
\be
e= e_0 + \Delta e \cos (gt) \ ,
\label{eq:evar}  
\ee
and
\be
\varpi = \varpi_0+ w\sin (gt) +g_2t \ .
\label{eq:pivar}  
\ee
As explained in section \ref{sec:sor}, we can neglect the effect of $g_2$ on the spin dynamics. The value of $\Delta e$ increases when the difference between the initial eccentricities of the orbits of the co-orbital bodies also increases \citep[see][]{Nauenberg2002}.
In addition, $\Delta e$ also increases when $\delta$ tends to 1 ($m_1 \ll m_2$).
We now head back to the general form of the Hamiltonian $\gH$ developed in section \ref{sec:sor}. By replacing the orbital variations given by expressions (\ref{eq:evar}) and (\ref{eq:pivar}) in equation (\ref{eq:H1gen}), we obtain a Hamiltonian that takes into account the variations of $e$ and $\varpi$. At first order in $w$ and $\Delta e$, following the method described in sections \ref{sec:ecckep} and \ref{sec:CoC}, we can split $\gH_1$ into a term which does not depend on $\lambda$ and a term that has zero as mean value over $\lambda$. 
After averaging, for $p=1$, $\overline{\gH}_{1,p}$ becomes:
\be
\begin{disarray} {ll}
\overline{\gH}_{1,p=1}=  - \frac{\sigma^2}{4} \cos  (2\gamma_1 - 2  \delta \hat{\zeta}(\tau)) \ ,
\end{disarray} 
\label{eq:resoeqevar}
\ee
while for $p=3/2$ or $1/2$ we have
\be
\begin{disarray} {ll}
\overline{\gH}_{1,p}= - \frac{\sigma^2}{4}&  \left[   X^{-3,2}_p(e_0)  \cos  (2\gamma_p - 2 p \delta \hat{\zeta}(\tau)  ) \frac{}{} \right.\\
 & \left. + X^{-3,2}_p(\frac{\Delta e + w e_0}{2})  \cos  (2\gamma_p - 2 p \delta \hat{\zeta}(\tau) + g t   ) \right. \ \\
 & \left. + X^{-3,2}_p(\frac{\Delta e - w e_0}{2})  \cos  (2\gamma_p - 2 p \delta \hat{\zeta}(\tau) - g t  ) \right] \ .
\end{disarray} 
\label{eq:resoeqevar2}
\ee

 \begin{figure}[h!]
\begin{center}
\hspace{0.3cm}
\includegraphics[scale=1]{\figpath/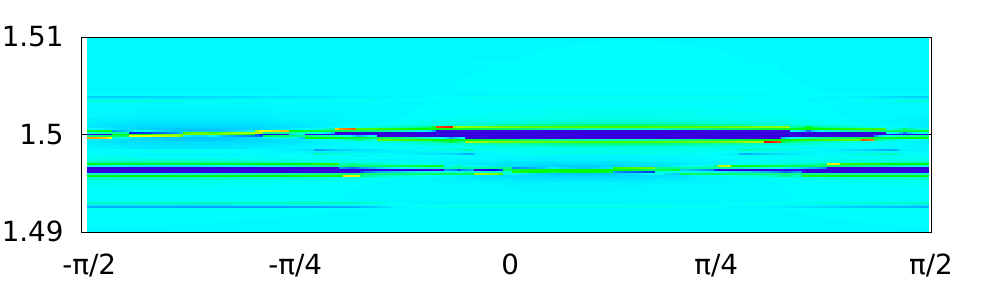}\\
\rotatebox{90}{\hspace{30pt} $\dot{\theta}/n $}
\includegraphics[scale=1]{\figpath/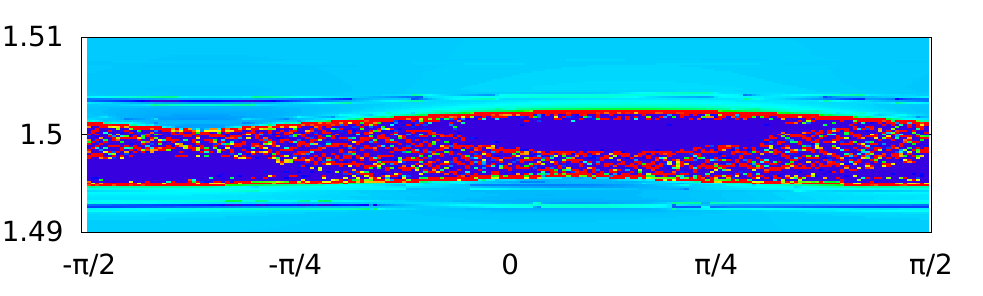}\\
\hspace{0.3cm}
\includegraphics[scale=1]{\figpath/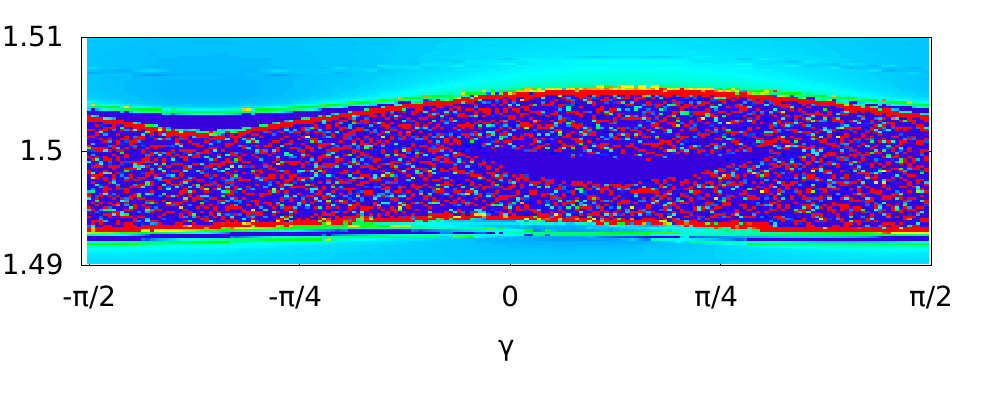}
\end{center}
\caption{\label{fig:echaos} FMA of the phase space ($\dot{\theta}/n,\gamma_p$) in the neighbourhood of the 3:2 spin orbit resonance, for two coorbitals on tadpole configuration. $e_1=0.05$, $e_2=0.1$, $m_1=m_2=1\times 10^{-3}m_0$ and $\zeta_0=45^\circ$. We have $\frac{B-A}{C}=1\times 10^{-6}$ on the top panel, $1\times 10^{-5}$ on the middle one and $5\times 10^{-5}$ on the bottom one. See the text for more details.}
\end{figure}

\begin{figure}[h!]
\begin{center}
\rotatebox{90}{\hspace{110pt} $\dot{\theta}/n$}
\includegraphics[scale=0.8]{\figpath/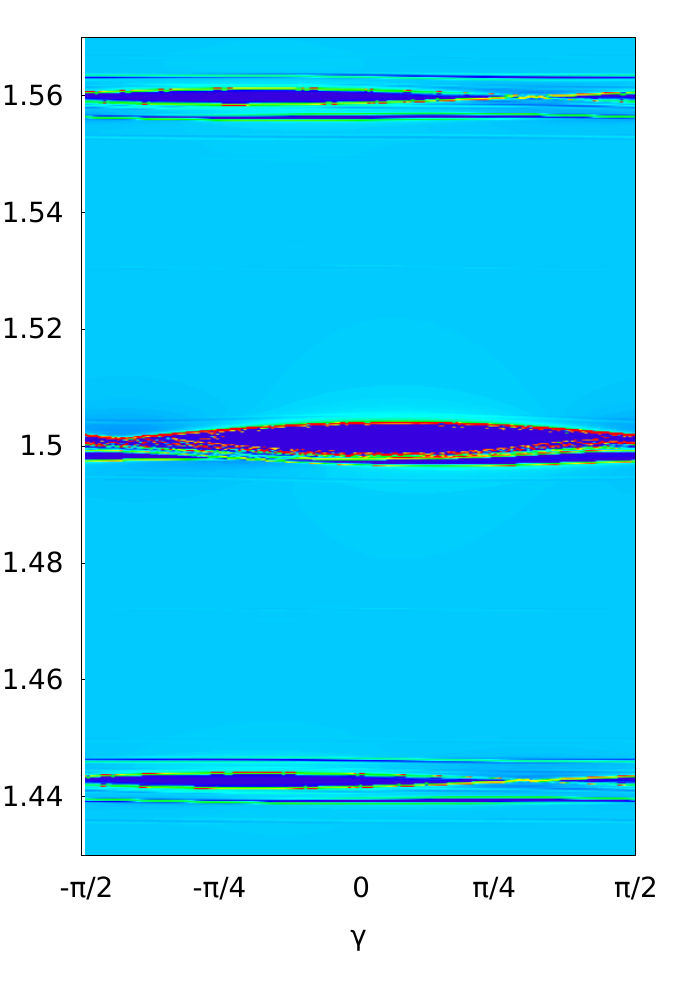}\includegraphics[scale=0.8]{\figpath/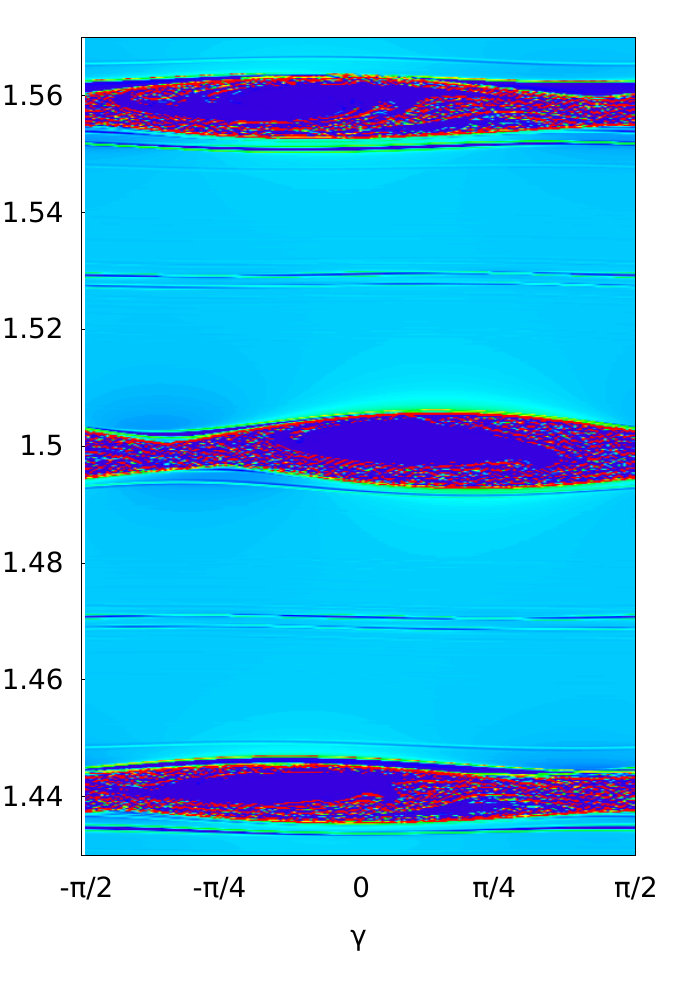}
\end{center}
\caption{\label{fig:fm32a1}FMA of the phase space ($\dot{\theta}/n,\gamma_p$) in the neighbourhood of the 3:2 spin orbit resonance, for two coorbitals on tadpole configuration. $e_1=e_2=0.1$, $\frac{B-A}{C}=5\times 10^{-5}$ and $\zeta_0=45^\circ$ in both cases. On the left $m_1=m_2=1\times 10^{-3}m_0$ and $m_2=2\times 10^{-3}m_0$, $m_1= 1\times 10^{-5}$  on the right. See the text for more details.}
\end{figure}

As in the cases previously studied, the rotational dynamics in the vicinity of the synchronous resonance remains unchanged by the eccentricity variations at first order in $\Delta e$ and $w$. 
However, the dynamics in the vicinity of the remaining eccentric spin-orbit resonances is modified by the variations of $e$ and $\varpi$.
We define $ \gamma_{p,\pm g}\equiv \gamma_p \pm gt $. 
Then, each term of the equation (\ref{eq:resoeqevar2}) can be rewritten under the form of the equation (\ref{eq:resoeq}). As a result, the $\rho_{\eta_j}$ obtained in section \ref{sec:SAS} can also be adapted to this case. As we saw previously, if $p=1$, only the $\rho_{(1,q,0)}$ are not equal to zero and their values have been given in section \ref{sec:SAS}. For $p=3/2$ or $1/2$, we get:
\be
\begin{disarray} {ll}
\rho_{(p,q,0)}&  =  \sigma \sqrt{X^{-3,2}_p(e_0)K^p_q(\delta,\zeta_0)},\\
\rho_{(p,q,\pm 1/2)}&  =  \sigma \sqrt{X^{-3,2}_p(\frac{\Delta e \pm w e_0}{2})K^p_q(\delta,\zeta_0)},
\end{disarray} 
\label{eq:rhoevar}
\ee
while the width of the higher order resonant islands is negligible within our approximation.  
The variations of $e$ and $\varpi$ thus split the eccentric spin orbit resonance and the co-orbital spin-orbit resonances into $3$ islands, whose center is separated by $g/2$ in the $\dot{\theta}$-direction in the phase space. This new effect on the phase space is more significant for massive co-orbitals. Indeed, we recall that $g=\gO(\mu)$, so the more massive the co-orbitals are, the larger the distance between the center of these islands, which increases their impact on the spin dynamics.  
The islands in $\pm g/2$ are not symmetric, the island in $+g/2$ is always bigger that the one in $-g/2$, which shrinks when $\Delta e \approx e_0w$. The asymmetry can be seen in the Figure~\ref{fig:echaos}: we force $e$ and $\varpi$ to oscillate by adopting different initial eccentricities $e_1=0.05$ and $e_2=0.1$ for $m_1$ and $m_2$, respectively. 
Focusing on the resonances located at $\eta_j=(3/2,0,s)$ with $s\in\{-1/2,0,1/2\} $ we can see that the islands located at $\pm g/2$ are asymmetric. As shown in equation (\ref{eq:rhoevar}), the island located at $\eta_j=(p,q,-1/2)$ can be of zero-width, while the island located in $\eta_j=(p,q,+1/2)$ can be of commensurate width with $\eta_j=(p,q,0)$. On the top panel, the islands at $\pm g/2$ are separated from the main island, while on the two others the three island overlap, ensuring chaos. 

In Figure~\ref{fig:fm32a1} we show another two examples where $g$ is large enough with respect to the island's width, such that it has an impact on the spin dynamics.
Here the initial eccentricities have been taken equal, so their oscillation is moderate and the islands in $s=0$ are sensibly larger than the ones in $s=\pm 1/2$. In both graphs the main $3$:$2$ island is located in $\dot{\theta}/n = 1.5$. The two coorbital spin-orbit resonances located at $\eta_j=(3/2,\pm 1/2,0)$ are easily identifiable at $\dot{\theta}/n = 1.5 \pm \nu/2 = 1.5 \pm 0.06$.
The split of these 3 resonances into 6 additional resonances centred at $\eta_j=(3/2,q,\pm1/2)$ can be seen on the left panel, where $\delta=1/2$: $\dot{\theta}/n = 1.5 \pm g/2 = 1.5 \pm 0.0036$ and $\dot{\theta}/n = 1.5 \pm \nu/2 \pm g/2$. 
On the right panel, the resonance width has been increased by taking a higher value of $\delta$ (increasing $(B-A)/C$ would have the same effect). We obtain a phase space that is similar to the one obtained when $e$ and $\varpi$ were considered constant (Fig.~\ref{fig:fm32a}). However, the overlap with the resonances in $\eta_j=(p,q,\pm 1/2)$ widens considerably the chaotic border of the islands \citep[][page 222]{2002M}.

\section{Conclusion}

The presence of a co-orbital companion leads to a perturbation of the true longitude at a frequency of order $\nu=\gO(\sqrt{\mu})$. This variation splits the eccentric spin-orbit resonances into new families of co-orbital spin-orbit resonances. Depending on the orbit, on the mass of the two orbiting bodies and on the axial asymmetry of the rotating body, those resonances can give rise to additional resonant islands, a chaotic region, or just widen the separatrix of the already existing islands.
  
Considering a system of two coorbital planets around a star, one can devise some interesting scenarios: for two super-earth with low asymmetry, their $\sigma$ is very small compared to $\nu$, thus their resonant islands are relatively thin and spaced from each other. 
On the contrary, for two earth-like planets, their $\sigma$ is large enough to a create wide chaotic area. 
For instance, for two Earth-like co-orbital planets, $\nu$ and $\sigma$ are commensurate ($\sigma/n \approx 7\times 10^{-3}$ and $\nu/(2n) \approx 3.5\times 10^{-3}$). 
Finally, if one of the planets is earth-like and the other is a gaseous giant, the phase space of the spin dynamics of the terrestrial planet consists in large separated resonant islands. 
More generally, due to the diversity of exoplanets and moons, any configuration is possible.

We have shown that co-orbital bodies with low libration amplitude of the resonant angle $\zeta$ (tadpole orbits) are likely trapped in a spin-orbit resonance located close to the classic eccentric resonances of the unperturbed problem \citep{1966GP}. 
As the libration amplitude of $\zeta$ increases, the co-orbital resonant islands become of commensurable width with the eccentric resonances. In the horseshoe domain, the eccentric spin-orbit resonances are of negligible width with respect to some of the co-orbital resonant island under and above them. In this configuration, the vicinity of the eccentric resonance is generally regular, while it can be surrounded by two chaotic areas created by the overlapping of the main co-orbital resonances.

The co-orbital bodies also mutually perturb the eccentricities and the argument of perihelion of each orbit on a long time-scale of order $\gO(\mu)$. 
The precession of the perihelion has no effect on the phase-space except a slight global offset in the direction of $\dot{\theta}$. 
However, the frequency $g$, which rules the libration of the perihelion and the oscillation of the eccentricity, splits both the eccentric spin-orbit families and the co-orbital spin-orbit families into a yet again new resonant family. 
Nevertheless, in this case the width of the resonant island depends on the variation of the eccentricity and on the argument of perihelion of the rotating body, while its separation to the eccentric or co-orbital spin-orbit resonance is $g/(2n) = \gO(\mu)$. 
Therefore, usually these islands have an insignificant width, or only slightly widen the separatrix of the main resonant islands. 
However, for systems with low axial asymmetry (low $\frac{B-A}{C}$) and massive co-orbital companions, such as gaseous giants, these resonances can have a significant impact on the spin dynamics.

\bibliographystyle{apalike}

\appendix

\section{Appendix}
\label{sec:bmasc}

\renewcommand{\arraystretch}{1.5}
\begin{table}[H!]
\begin{center}
\caption{Notations. All orbital elements are given in heliocentric coordinates, thus with respect to $m_0$}
\label{tab:parf}
\begin{tabular}{|c|c|}
\hline
variable & definition \\ 
\hline
  $a_j$ & semi-major axis of body $j$ \\
$\lambda_j$ & mean longitude of body $j$ \\
   $e_j$ & eccentricity of body $j$ \\  
   $\varpi_j$ & argument of periastron of body $j$ \\ 
    $m_j$ & mass of body $j$ \\ 
    $r_j$ & distance between $m_0$ and $m_j$ \\
    $f_i$ & true anomaly of body $j$ \\
 $\bar{a}$ & mean semi-major axis of the co-orbital motion \\      
 $n$ & mean mean motion, defined from $\bar{a}$ \\
 $\mu$ & $= (m_1+m_2)/(m_0+m_1+m_2)$ \\
   $\delta$ & $= m_2/(m_1+m_2)$ \\
  $\zeta$ & $= \lambda_1-\lambda_2$ semi-fast resonant angle\\
 $\zeta_0$ & minimum value of $\zeta$ on a given orbit \\
  $\zeta_s$ & value of $\zeta$ at the separatrix between tadpole and horseshoe  \\
   $\zeta_L$ & value of $\zeta$ at the considered Lagrangian equilibrium  \\
   $z$ & $=\zeta_0-\zeta_L$ \\
   $\hat{\zeta}$ & such that $\hat{\zeta}(\tau)=\zeta(\tau/\nu) $\\
 $\nu$ & fundamental frequency for the evolution of $\zeta$ \\
 $g_1$ and $g_2$ & fundamental frequencies for the evolution of ($\varpi_1$,$\varpi_2$) \\
  $g$ & $= g_1- g_2$ \\
    $\Delta \varpi$ & $= \varpi_1-\varpi_2$ slow resonant angle\\
  $\theta$ & rotation angle of $m_1$ \\
  $A$, $B$, $C$ & moment of inertia of $m_1$ \\
  $\sigma$ & $=n \sqrt{3(B-A)/C}$ asymmetry of the rotating body\\
  $\varsigma$ & $=(n-g_2,\nu,g) \in  \mathbb{R}^3_+$ frequency vector \\
  $\eta_j$& $=(p,q,s)$ with $2\eta_j \in \mathbb{Z}^3$ identification of a given resonance \\
  $\rho_{\eta_j}$ & width of the resonance  $\eta_j$ \\
  $\gamma_p$ & $\theta - pnt$ \\
 $ \epsilon^k_j$ & $=|\langle \eta_j-\eta_k, \varsigma \rangle|$ \\
 $X^{-3,2}_p(e)$  & Hansen coefficient defined eq. (\ref{eq:hansen}) \\
 $\xi_l$  & coefficient of the Fourier expansion of $\hat{\zeta}$  \\
  $J_k$  & $k^{th}$ Bessel function  \\
 \hline
\end{tabular}
\end{center}
\end{table}

\begin{table}[H!]
\begin{center}
\caption{$(B-A)/C$ for bodies of the solar system \citep[see][and references therein]{CoRod2013,RoRaEl2012}. The $(B-A)/C$ of the co-orbitals orbiting around Saturn are so large that their resonant islands totally overlap \citep{RoRaEl2012}.}
\label{tab:parf2}
\begin{tabular}{|c|c|c|}
\hline
 body & (B-A)/C & status \\ 
\hline
Mercury & $ 8.1 \times 10^{-5}$ & planet\\
Venus &  $5.39 \times 10^{-6}$ & planet\\
Earth & $ 1.57 \times 10^{-5}$ & planet\\
Mars & $ 5.5 \times 10^{-4}$ & planet\\
Moon & $ 2.2 \times 10^{-4}$ & satellite\\
Io & $ 5.6 \times 10^{-3}$ & satellite\\
Europa & $1.3 \times 10^{-3}$ & satellite\\
Ganymede & $ 3.8 \times 10^{-4}$ & satellite\\
Callisto & $1.0 \times 10^{-4}$ & satellite\\
Rhea & $2.4 \times 10^{-3}$ & satellite\\
Titan & $1.0\times 10^{-4}$ & satellite\\
Polydeuces & $2.2 \times 10^{-1}$ & satellite (tadpole)\\
Helene & $1.3 \times 10^{-1}$ & satellite (tadpole) \\
Telesto & $3.2 \times 10^{-1}$ & satellite (tadpole) \\
Calypso & $2.7 \times 10^{-1}$ & satellite (tadpole) \\
Janus & $1.0 \times 10^{-1}$ & satellite (horseshoe) \\
Epimetheus & $3.0 \times 10^{-1}$ & satellite (horseshoe) \\
\hline
\end{tabular}
\end{center}
\end{table}


\end{document}